\documentclass[proceedings]{JHEP3}
\usepackage{eurosym}
\usepackage{amssymb}
\usepackage{amsfonts}
\usepackage{amsmath,epsfig}
\usepackage{graphicx}
\usepackage{xcolor}

\newbox\mybox

\newcommand\fverb{\setbox\mybox=\hbox\bgroup\verb}
\newcommand\fverbdo{\egroup\medskip\noindent\fbox{\unhbox\mybox}\ }
\newcommand\fverbit{\egroup\item[\fbox{\unhbox\mybox}]}
\conference{Massive gauge particles versus Goldstone bosons }
\abstract{We investigate the Englert-Brout-Higgs-Guralnik-Hagen-Kibble mechanism for non-Hermitian field theories with local non-Abelian gauge symmetry in
different regions of their parameter spaces. We demonstrate that the two aspects of the mechanism, that is giving mass to gauge vector bosons and at the same time preventing the
existence of massless Goldstone bosons, remain to be synchronized in all regimes characterized by a modified CPT symmetry. 
In the domain of parameter space where the ``would be Goldstone bosons'' can be identified the gauge vector bosons become massive and the Goldstone bosons cease to exist. 
The mechanism is also in tact at the standard exceptional points. However, at the zero exceptional points, that is when the eigenvalues of the mass squared matrix vanish 
irrespective of the symmetry breaking, the mechanism breaks down as the Goldstone bosons can not be identified 
and the gauge vector bosons remain massless. This breakdown coincides with the vanishing of the CPT inner product of symmetry breaking vacua defined on the eigenvector 
space of mass squared matrix.
We verify this behaviour for a theory with SU(2) symmetry in which the complex scalar fields are taken 
in the fundamental as well as in the adjoint representation.}

\title{Massive gauge particles versus Goldstone bosons in non-Hermitian
non-Abelian gauge theory}
\author{Andreas Fring and Takanobu Taira \\
Department of Mathematics, City, University of London,\\
Northampton Square, London EC1V 0HB, UK \\
E-mail: a.fring@city.ac.uk, takanobu.taira@city.ac.uk}

\input{tcilatex}
\begin{document}

\section{Introduction}

Our main objective is to extend the
Englert-Brout-Higgs-Guralnik-Hagen-Kibble mechanism \cite{englert1964broken,higgs1964brokena,higgs1964brokenb,guralnik1964global}, hereafter simply referred to as Higgs mechanism, to non-Hermitian field
theories with a local non-Abelian gauge symmetry using a pseudo-Hermitian
approach. We focus on the two key aspects for which the mechanism was
originally developed, that is to give mass to gauge vector bosons and at the
same time prevent the existence of massless Goldstone bosons. When keeping
the symmetry global one may adopt different starting points for the study of
Goldstone phases, such as the field content of local operators, a scattering
matrix based on a particle picture or an explicit Lagrangian.

For instance, two dimensional conformal quantum field theories are well
understood in terms of their operator content characterized by
infinite-dimensional algebras of local conformal transformations \cite%
{Goddard:1986bp}. A large class of such theories, minimal models \cite{BPZ},
are known to possess a finite operator content and the treatment of unitary
and non-unitary theories is formally identical. The simplest massive
non-unitary field theory consisting of only one real scalar field describing
in its ultraviolet limit the critical point of the Ising model in a purely
imaginary magnetic field, the Yang-Lee edge singularity \cite{YL1,YL2}, is
known for a long time to correspond to the non-Hermitian Lagrangian \cite%
{fisher1978yang,cardy1985conformal}%
\begin{equation}
	\mathcal{L}=\int d^{d}x\left[ \frac{1}{2}\left( \bigtriangledown \phi
	\right) ^{2}+i(h-h_{c})\phi +\frac{1}{3}ig\phi ^{3}\right] .  \label{phi3}
\end{equation}%
Exact scattering theories for two-dimensional models have also been
identified \cite{cardy1989s}, that can be used to probe the ultraviolet
limit most easily by employing the thermodynamic Bethe Ansatz \cite%
{zamothermo}. These techniques have also been employed for hypothetical
scattering matrices for massless Goldstone fermions (Goldstinos)\ \cite%
{zamo1991tri} and scattering matrices that reduce to them in certain limits 
\cite{TBAHSG1999}. Despite the fact that the Mermin-Wagner theorem prevents
the validity of the Goldstone theorem in dimensions $d\leq 2$, it was argued
in \cite{jacobsen2003} that for certain symmetry groups, e.g. $SO(N)$ with
formally taking $N<2$, this restriction can be circumvented so that
Goldstone phases maybe be identified in such type of non-Hermitian systems.

Rather than taking an operator content or a scattering matrix as a starting
point, one may of course also commence directly with a non-Hermitian
Lagrangian. From that perspective it is natural to try to extend techniques
and methods developed for the treatment of non-Hermitian quantum mechanics 
\cite{Geyer,Bender:1998ke,Alirev,PTbook} to a quantum field theory setting.
Such considerations have been carried out for a scalar field theory with
imaginary cubic self-interaction terms \cite{benderphi31,shalabyphi31}, with
a Lagrangian identical to (\ref{phi3}) but for $h=h_{c}$ without a linear
term, deformed harmonic oscillators \cite{bender2018p}, non-Hermitian
versions with a field theoretic Yukawa interaction \cite%
{alexandre2015non,rochev2015hermitian,korchin2016Yuk,laureyukawa}, free
fermion theories with a $\gamma _{5}$-mass term and the massive Thirring
model \cite{bender2005dual}, $\mathcal{PT}$-symmetric versions of quantum
electrodynamics \cite{bender1999nonunitary,bender2000solution,milton2013pt}
and other types of $\mathcal{PT}$-symmetric quantum field theories in higher
dimensions \cite{bender2018p} than (\ref{phi3}).

Non-Hermitian quantum field theories are also viewed as possible options to
overcome the limitations of the Standard Model of particle physics, which is
well known to fail in describing gravity and neutrino oscillations in a
consistent manner. In \cite{Ohlsson} a study was initiated to potentially
resolving the latter issue by extending the ordinary two-flavor neutrino
oscillation to a non-Hermitian PT-symmetric/pseudo Hermitian setting. A well
studies simple Hermitian extension of the Standard Model referred to as the
two-Higgs-doublet-model \cite{twoHiggs} involves a second Higgs doublet or
possibly more \cite{multiHiggs}. Here we investigate non-Hermitian variants
of these models.

Here we are especially interested in complex non-Hermitian scalar field
theories and the question of how the aforementioned Goldstone phases
manifest in these theories, together with the subsequent extension to the
Higgs mechanism \cite%
{englert1964broken,higgs1964brokena,higgs1964brokenb,guralnik1964global} in
Abelian and non-Abelian gauge theories. These issues have been studied
recently by various groups in different approaches, which differ from their
very onset: Given a generic action for a complex scalar field theory of the
form $\mathcal{I}=\int d^{4}x\mathcal{L}(\phi ,\phi ^{\ast })$, one has two
options in a Hermitian theory to derive the equations of motion by means
functional variation, either to calculate $\delta \mathcal{I}/\delta \phi =0$
or $\delta \mathcal{I}/\delta \phi ^{\ast }=0$. Since the standard $\mathcal{%
CPT}$-theorem \cite{schwinger51} applies, the two resulting equations are
the same. In contrast, in a non-Hermitian theory one no longer has $\mathcal{%
I}=\mathcal{I}^{\ast }$, so that the two equations are not only not the
same, but in addition one also has the new options $\delta \mathcal{I}^{\ast
}/\delta \phi =0$ and $\delta \mathcal{I}^{\ast }/\delta \phi ^{\ast }=0$.
In the first approach, we refer to as the "\emph{surface term approach}", it
was suggested \cite{alexandre2018spontaneous,alex2019} to take of only the
two equations resulting from $\delta \mathcal{I}/\delta \phi =0$, $\delta 
\mathcal{I}^{\ast }/\delta \phi ^{\ast }=0$ and neglect the remaining two.
As the resulting equations are in general not compatible, the authors
propose to use some non-vanishing surface terms to compensate for the
discrepancy. The second approach \cite{mannheim2018goldstone} consists of
taking $\delta \mathcal{I}/\delta \phi =0$ or $\delta \mathcal{I}/\delta
\phi ^{\ast }=0$, and when determining the vacuum allowing the real and imaginary parts of the complex scalar field to be also complex. Thus in this approach the field content is doubled or re-defined.

Here we follow an approach, we refer to as the "\emph{pseudo-Hermitian
approach}" \cite{fring2019pseudo,fring2020goldstone}, more aligned to the
procedure pursued in non-Hermitian versions of quantum mechanics, in which
one employs so-called Dyson maps \cite{Dyson} to transform a non-Hermitian
Hamiltonian to a Hermitian Hamiltonian. Since the action $\mathcal{I}$
contains a Lagrangian, rather than a Hamiltonian, we need to first Legendre
transform the complex Lagrangian $\mathcal{L}$ to a non-Hermitian
Hamiltonian $\mathcal{H}$, carry out the similarity transformation by means
of a Dyson map, while preserving equal time commutation relations, to obtain
a Hermitian Hamiltonian $\mathfrak{h}$, which we then inverse Legendre
transform to a real Lagrangian $\mathfrak{l}$ 
\begin{equation}
\mathcal{L}\overset{\text{Legendre}}{\rightarrow }\mathcal{H}\overset{\text{%
Dyson}}{\rightarrow }\eta \mathcal{H}\eta ^{-1}=\mathfrak{h}\overset{\text{%
Legendre}^{-1}}{\rightarrow }\mathfrak{l.}  \label{1}
\end{equation}%
A consistent set of equations of motion is then obtained by functionally
varying the action $\mathfrak{s}=\int d^{4}x\mathfrak{l}(\varphi ,\chi )$
involving this real Lagrangian $\mathfrak{l}$ with respect to the real field
components $\varphi $, $\chi \in \mathbb{R}$ of the complex scalar field $%
\phi =1/\sqrt{2}(\varphi +i\chi )$, i.e. $\delta \mathfrak{s}/\delta \varphi
=0$ and $\delta \mathfrak{s}/\delta \chi =0$. In order to perform the
similarity transformation, one needs to canonically quantize the theory
first,  which is what we will do below. 

In fact, this will be the only quantum aspect in our manuscript. Our discussion here is kept classical and we will not carry out a second quantization of the scalar fields, nor BRST quantization for the gauge fields, etc. Supported by our analysis below and in our previous papers
 \cite{fring2019pseudo,fring2020goldstone}, our view here is that much insight can be gained by a classical treatment, apart from the appeal to the quantum field theory for the canonical commutation relations. One may of course fully analyse the theory in its quantum aspects, but our view is that this should be performed on the transformed, i.e. Hermitian, theory rather than directly on the non-Hermitian theory. In the former scenario one would not expect any issues to arise as one is simply dealing with a Hermitian system, whereas in the latter case one may encounter a number of what we would refer to as ``\textit{pseudo-problems}''. 
 	
 	The above mentioned incompatibility of the variational principle is an example for such a problem that already occurs on the level of the classical theory. One may try to fix this problem directly for the non-Hermitian  system or simply consider the equivalent Hermitian system in which the problem is entirely absent. Another
 well known example for a pseudo-problem in the quantum mechanical context is for instance
 to take the variables $x$ and $p$, that might appear in the definition of  a non-Hermitian Hamiltonian, to be physical observables. However, only the pseudo-Hermitian counterparts or the Dyson mapped quantities $\eta x \eta^{-1}$ and $\eta p \eta^{-1}$ can be observed and should be interpreted as being physical. Taking instead the original operators $x$ and $p$ from the non-Hermitian theory one may derive violations of the uncertainly relations and other properties that contradict standard quantum mechanical principles. Again these problems entirely disappear in the equivalent Hermitian system when considering the correct variables.    

Previously we have analysed the Goldstone theorem for a non-Hermitian scalar field theory with an Abelian \cite{fring2019pseudo} and a non-Abelian symmetry \cite{fring2020goldstone}. Here the main purpose is to investigate the extension to the Higgs mechanism. As it is well known, in the standard Higgs mechanism the Goldstone bosons acquires a mass so that our previous findings will inevitably have a bearing on the investigation to be carried out here. Let us therefore recall the key finding from \cite{fring2019pseudo,fring2020goldstone}: The main object of study has been the eigenvalue spectra of the
non-Hermitian squared mass matrix $M^{2}$, obtained by expanding around the
symmetry breaking or preserving vacua. The reality of these eigenvalues has been guaranteed by a modified $\mathcal{CPT}$ -symmetry of the original
action $\mathcal{I}$. Hence, we distinguished in the usual fashion between $%
\mathcal{CPT}$\emph{-symmetry regime} characterized by $M^{2}$ commuting
with this symmetry operator and its eigenstates being simultaneous
eigenstates of the symmetry operator. When the latter is not the case, one
refers to that regime as the $\mathcal{CPT}$\emph{-spontaneously broken
regime} in which some eigenvalues become complex conjugate pairs. The points
in parameter space at which this occurs are commonly referred to as \emph{%
exceptional point}. As physical masses are positive and real, we also
require the eigenvalues of $M^{2}$ to be non-negative. We encountered a
special behaviour at the transition points when the eigenvalues become zero,
which we referred to \cite{fring2019pseudo,fring2020goldstone} as \emph{zero
exceptional points of type I and type II}. \footnote{The difference between these two types of points is discussed in the appendix in form of a general discussion for key blocks of the squared mass matrix.} At the type I points the mass
matrix is non-diagonalizable and the continuous symmetry is broken, whereas
at the type II points the mass matrix can be diagonalized and the vacuum
with broken continuous symmetry re-acquires the symmetry at this point.

Using the above mentioned approaches, the Higgs mechanism was previously
studied for Abelian \cite{mannheim2018goldstone,alexanHiggs} as well as
non-Abelian gauge theories \cite{alex2019} leading to slightly different
findings. In \cite{mannheim2018goldstone} the interesting observation made,
that the mass of the gauge vector boson vanishes at the zero exceptional
point, was not confirmed in \cite{alexanHiggs}. In addition, for the
non-Abelian gauge theories it was found in \cite{alex2019} that the Higgs
mechanism even applies in the spontaneously broken $\mathcal{CPT}$ -regime.
Our aim is here to compare the various observations made using these
alternative approaches with a pseudo-Hermitian approach, extend the studies
to other models, symmetries and representations within this framework.

Our manuscript is organised as follows: In section 2.1 we introduce first our non-Hermitian model with scalar field taken in the fundamental representation possessing a local  $%
	SU(2)\times U(1)$-symmetry, discuss the symmetry breaking vacuum of this theory and subsequently the Higgs mechanism. We indicate how to extend the model from $SU(2)$ to $SU(N)$. We repeat the same discussion in section 2.2 for anon-Hermitian model with scalar field taken in the adjoint representation. Our conclusions are stated in section 3. The distinction between the two types of exceptional points is crucial and for that reason we include in our appendix a discussion for part of the squared mass matrix that explains the difference.

\section{Pseudo-Hermitian approach to the Higgs mechanism}

In this section we commence by investigating the same model considered in 
\cite{alex2019} using, however, a pseudo-Hermitian method to compare our
results with the findings in \cite{alex2019}. We will observe that the mass
spectrum of the fields in the $SU(2)$ fundamental representation coincides
with the one found in \cite{alex2019}, but the masses for the gauge vector
bosons differ and in particular vanish at the zero-exceptional points. We
will extend this model to incorporate a $SU(N)$-symmetry and continue to
observe this phenomena also for these more general systems. Finally we will
consider a new model for which the fields are taken in a different
representation, the adjoint representation of $SU(2)$, making similar
observations.

\subsection{A $SU(2)$-model in the fundamental representation}

We start by applying the pseudo-Hermitian approach to a model with local $%
SU(2)\times U(1)$-symmetry previously studied using the surface term
approach in \cite{alex2019}. The model corresponds to the gauged version of
the one for which the Goldstone mechanism was studied in \cite%
{fring2019pseudo} 
\begin{equation}
\mathcal{L}_{2}=\sum_{i=1}^{2}\left\vert D_{\mu }\phi _{i}\right\vert
^{2}+m_{i}^{2}|\phi _{i}|^{2}-\mu ^{2}(\phi _{1}^{\dagger }\phi _{2}-\phi
_{2}^{\dagger }\phi _{1})-\frac{g}{4}(|\phi _{1}|^{2})^{2}-\frac{1}{4}%
\limfunc{Tr}\left( F_{\mu \nu }F^{\mu \nu }\right) . \label{2.1}
\end{equation}%
Here $g,\mu \in \mathbb{R}$, $m_{i}\in \mathbb{R}$ or $m_{i}\in i\mathbb{R}$
are constants. When compared to \cite{fring2019pseudo} we have replaced here
as usual the standard derivatives $\partial _{\mu }$ by covariant
derivatives $D_{\mu }:=\partial _{\mu }-ieA_{\mu }$, involving a charge $%
e\in \mathbb{R}$ and the Lie algebra valued gauge fields $A_{\mu }:=\tau
^{a}A_{\mu }^{a}$. Here the $\tau ^{a}$, $a=1,2,3$, are taken to be Pauli
matrices, which when re-defined as $i(-1)^{a+1}\tau ^{a}$ are the generators
of $SU(2)$. We have also added the standard Yang-Mills term comprised of the
Lie algebra valued field strength $F_{\mu \nu }:=\partial _{\mu }A_{\nu
}-\partial _{\nu }A_{\mu }-ie[A_{\mu },A_{\nu }]$. The two complex scalar
fields $\phi _{i}$ are taken to be in the representation space of
fundamental representation of $SU(2)$. The model described by $\mathcal{L}%
_{2}$ admits a global continuous $U(1)$-symmetry, a local continuous $SU(2)$%
-symmetry and two discrete antilinear $\mathcal{CPT}$-symmetries as
described in more detail in \cite{fring2019pseudo}. Crucially $\mathcal{L}%
_{2}$ is not Hermitian, which at this point is simply to be understood as
not being invariant under complex conjugation. The Abelian version of $%
\mathcal{L}_{2}$ was discussed in \cite%
{mannheim2018goldstone,fring2020goldstone}.

As argued in \cite{fring2019pseudo}, it is useful to decompose the complex
fields into their real components $\phi _{j}^{k}=1/\sqrt{2}(\varphi
_{j}^{k}+i\chi _{j}^{k})$ with $\varphi _{j}^{k}$, $\chi _{j}^{k}\in \mathbb{%
R}$.  Thus simply rewriting the complex scalar fields in  equation (2.1) in terms of their real and imaginary components we obtain the following Lagrangian 
\begin{eqnarray}
&&\mathcal{L}_{2}=\frac{1}{2}\sum_{k=1}^{2}\sum_{j=1}^{2}\left\{ \left[
\partial _{\mu }\varphi _{j}^{k}+e(A_{\mu }\chi _{j})^{k}\right] \left[
\partial ^{\mu }\varphi _{j}^{k}+e(A^{\mu }\chi _{j})^{k}\right] ^{\ast
}\right.  \label{LN} \\
&&+\left[ \partial _{\mu }\chi _{j}^{k}-e(A_{\mu }\varphi _{j})^{k}\right] %
\left[ \partial ^{\mu }\chi _{j}^{k}-e(A^{\mu }\varphi _{j})^{k}\right]
^{\ast }\!\!\!-2\!\func{Im}\left[ \left[ \partial _{\mu }\varphi
_{j}^{k}+e(A_{\mu }\chi _{j})^{k}\right] ^{\ast }\!\left[ \partial ^{\mu
}\chi _{j}^{k}-e(A^{\mu }\varphi _{j})^{k}\right] \right]  \notag \\
&&+\left. m_{j}^{2}\left[ (\varphi _{j}^{k})^{2}+(\chi _{j}^{k})^{2}\right]
-2i\mu ^{2}(\varphi _{1}^{k}\chi _{2}^{k}-\chi _{1}^{k}\varphi _{2}^{k})-%
\frac{1}{4}F_{\mu \nu }^{k}\left( F^{k}\right) ^{\mu \nu }\right\} -\frac{g}{%
16}\left[ \sum_{k=1}^{2}(\varphi _{1}^{k})^{2}+(\chi _{1}^{k})^{2}\right]
^{2} . \notag
\end{eqnarray}%
We use here the standard notation * for complex conjugation and $\dagger$ for the simultaneous conjugation with transposition. 
	
	The $SU(2)$-symmetry manifests itself as follows: A change in the complex scalar
	fields due to this symmetry is $\delta \phi _{j}^{k}=i\alpha _{a}T_{a}^{kl}\phi _{j}^{l}$, where
	the generators $T_{a}$~of the symmetry transformation are the standard
	Pauli matrices $\sigma _{a}$, $a=1,2,3$. The
	infinitessimal changes for the real component fields are then identified as
	\begin{eqnarray}
		\delta \varphi _{j}^{1} &=&-\alpha _{1}\chi _{j}^{2}+\alpha _{2}\varphi
		_{j}^{2}-\alpha _{3}\chi _{j}^{1},\quad \delta \chi _{j}^{1}=\alpha
		_{1}\varphi _{j}^{2}+\alpha _{2}\chi _{j}^{2}+\alpha _{3}\varphi _{j}^{1},
		\label{t1} \\
		\delta \varphi _{j}^{2} &=&-\alpha _{1}\chi _{j}^{1}-\alpha _{2}\varphi
		_{j}^{1}+\alpha _{3}\chi _{j}^{2},\quad \delta \chi _{j}^{2}=\alpha
		_{1}\varphi _{j}^{1}-\alpha _{2}\chi _{j}^{1}-\alpha _{3}\varphi _{j}^{2},
		\label{t2}
	\end{eqnarray}
	which leave the above Lagrangian invariant. 
	The discrete antilinear $\mathcal{CPT}_{\pm }$-symmetries manifest themselves as 
	\begin{equation}
		\mathcal{CPT}_{\pm } :~\varphi _{j}^{k}(x_{\mu })\rightarrow \mp
		(-1)^{j}\varphi _{j}^{k}(-x_{\mu }),~~\chi _{j}^{k}(x_{\mu })\rightarrow \pm
		(-1)^{j}\chi _{j}^{k}(-x_{\mu }).
	\end{equation}

	A noteworthy remark is that it is straightforward to generalise the model from a locally $
	SU(2) $-invariant one to a locally $SU(N)$-invariant one, by extending the sum over $k$ from $2$ to $N$, while keeping the $U(1)$-symmetry global. In what follow we will focus on $N=2$. 

A crucial feature of $\mathcal{L}_{2}$ is that its $\mathcal{CPT}$%
-invariance translates into pseudo-Hermiticity \cite%
{Mostafazadeh:2002pd,Alirev}, meaning that it can be mapped to a Hermitian
Lagrangian $\mathfrak{l}_{2}$ by means of the adjoint action of a Dyson map $%
\eta $ as $\mathfrak{l}_{{\color{red}2}}=\eta \mathcal{L}_{2}\eta ^{-1}$. This may be
achieved by the slightly modified version of the Dyson map used in \cite%
{mannheim2018goldstone,fring2019pseudo} 
\begin{equation}
\eta _{2}^{\pm }=\exp \left( \pm \sum_{i=1}^{2}\int d^{3}x\Pi ^{\varphi
_{2}^{i}}(t^{\prime },\Vec{x})\varphi _{2}^{i}(t^{\prime },\Vec{x})+\Pi
^{\chi _{2}^{i}}(t^{\prime },\Vec{x})\chi _{2}^{i}(t^{\prime },\Vec{x}%
)\right) .
\end{equation}%
We denote here the time-dependence by $t^{\prime }$ to indicate that
commutators are understood as equal time commutators  for 
the canonical momenta $\Pi^{\varphi_2^i }=\partial _{t}\varphi
_{2}^i$ and $\Pi^{\chi_2^i }=\partial _{t}\chi _{2}^i$, $i=1,2$. satisfying $\left[ \psi _{j}^k(\mathbf{x},t),\Pi^{\psi _{l}^m}(\mathbf{y},t)\right] =i \delta_{jl} \delta_{km} \delta (\mathbf{x}-\mathbf{y})$, $%
j,k,l,m=1,2$, for $\psi =\varphi ,\chi $.

Hence $\eta _{2}^{\pm
} $ is not to be viewed as explicitly time-dependent as discussed in much
detail for instance in \cite{frith2019time}. The adjoint action of $\eta
_{2}^{+}$ on the individual fields maps as 
\begin{equation}
\varphi _{1}^{k}\rightarrow \varphi _{1}^{k}~,~~\varphi _{2}^{k}\rightarrow
-i\varphi _{2}^{k}~,~~\chi _{1}^{k}\rightarrow \chi _{1}^{k}~,~~\chi
_{2}^{k}\rightarrow -i\chi _{2}^{k}~,~~A_{\mu }\rightarrow A_{\mu
}~,~~~~k=1,2.
\end{equation}%
Thus, we convert the complex Lagrangian into the real Lagrangian%
\begin{eqnarray}
&&\mathfrak{l}_{2}=\frac{1}{2}\sum_{j=1}^{2}(-1)^{j+1}\left\{ \left\vert
\partial _{\mu }\varphi _{j}+e(A_{\mu }\chi _{j})\right\vert ^{2}+\left\vert
\partial _{\mu }\chi _{j}-e(A_{\mu }\varphi _{j})\right\vert ^{2}+m_{j}^{2}%
\left[ \varphi _{j}\cdot \varphi _{j}+\chi _{j}\cdot \chi _{j}\right]
\right. ~~~~~~ \\
&&-\left. 2\!\func{Im}\left[ \left[ \partial _{\mu }\varphi _{j}+e(A_{\mu
}\chi _{j})\right] ^{\ast }\!\cdot \left[ \partial ^{\mu }\chi _{j}-e(A^{\mu
}\varphi _{j})\right] \right] +(-1)^{j}2\mu ^{2}(\varphi _{1}\cdot \chi
_{2}-\chi _{1}\cdot \varphi _{2})\right\}  \notag \\
&&-\frac{g}{16}\left[ \varphi _{1}\cdot \varphi _{1}+\chi _{1}\cdot \chi _{1}%
\right] ^{2}-\frac{1}{4}\limfunc{Tr}\left( F_{\mu \nu }F^{\mu \nu }\right) .
\notag
\end{eqnarray}%
We may transform here directly the Lagrangian rather than the Hamiltonian,
as suggested in (\ref{1}), since the kinetic energy term is real and the
complexity only result from the potential term. Introducing the $2$
two-component fields of the form 
\begin{equation}
\Phi ^{k}:=\left( 
\begin{array}{c}
\varphi _{1}^{k} \\ 
\chi _{2}^{k}%
\end{array}%
\right) ~,~~\Psi ^{k}:=\left( 
\begin{array}{c}
\chi _{1}^{k} \\ 
\varphi _{2}^{k}%
\end{array}%
\right) ,~~~\ ~~~k=1,2,
\end{equation}%
we can re-write the Lagrangians $\mathcal{L}_{2}$ and $\mathfrak{l}_{2}$
more compactly. Defining the $2\times 2$ matrices 
\begin{equation}
H_{\pm }:=\left( 
\begin{array}{cc}
-m_{1}^{2} & \pm \mu ^{2} \\ 
\pm \mu ^{2} & m_{2}^{2}%
\end{array}%
\right) ~,~~\mathcal{I}:=\left( 
\begin{array}{cc}
1 & 0 \\ 
0 & -1%
\end{array}%
\right) ~,~~E:=\left( 
\begin{array}{cc}
1 & 0 \\ 
0 & 0%
\end{array}%
\right) ,  \label{HIE}
\end{equation}%
the real Lagrangian $\mathfrak{l}_{2}$ acquires the form%
\begin{eqnarray}
\mathfrak{l}_{2} &=&\frac{1}{2}\left\{ \left[ \partial _{\mu }\Phi +e%
\mathcal{I}A_{\mu }\Psi \right] ^{\ast }\mathcal{I}\left[ \partial ^{\mu
}\Phi +e\mathcal{I}A^{\mu }\Psi \right] +\left[ \partial _{\mu }\Psi -e%
\mathcal{I}A_{\mu }\Phi \right] ^{\ast }\mathcal{I}\left[ \partial ^{\mu
}\Psi -e\mathcal{I}A^{\mu }\Phi \right] \right. \\
&&-2\func{Im}\left[ \left[ \partial _{\mu }\Phi +e\mathcal{I}A_{\mu }\Psi %
\right] ^{\ast }\left[ \partial ^{\mu }\Psi -e\mathcal{I}A^{\mu }\Phi \right]
\right] \left. -{\Phi }^{T}H_{+}\Phi -{\Psi }^{T}H_{-}\Psi \right\}  \notag
\\
&&-\frac{g}{16}\left( {\Phi }^{T}E\Phi +{\Psi }^{T}E\Psi \right) ^{2}-\frac{1%
}{4}\limfunc{Tr}\left( F_{\mu \nu }F^{\mu \nu }\right) .  \notag
\end{eqnarray}%
We have simplified here the index notation by implicitly contracting,
keeping in mind that we are summing over two separate index sets $k\in
\{1,2\}$ and $j\in \{1,2\}$. For instance, we set  
\begin{eqnarray}
(IA_{\mu }\Phi )_{\alpha }^{k} &\rightarrow &\mathcal{I}_{\alpha \beta
}A_{\mu }^{kj}\Phi _{\beta }^{j}, \\
\left[ \partial _{\mu }\Phi _{j}^{k}+e\left( \mathcal{I}A_{\mu }\Psi \right)
_{j}^{k}\right] ^{\ast }\mathcal{I}_{j\ell }\left[ \partial ^{\mu }\Phi
_{\ell }^{k}+e\left( \mathcal{I}A^{\mu }\Psi \right) _{\ell }^{k}\right]
&\rightarrow &\left[ \partial _{\mu }\Phi +e\mathcal{I}A_{\mu }\Psi \right]
^{\ast }\mathcal{I}\left[ \partial ^{\mu }\Phi +e\mathcal{I}A^{\mu }\Psi %
\right] ,~~~~~~~ \\
{\Phi ^{k}}^{T}H_{+}\Phi ^{k} &\rightarrow &{\Phi }^{T}H_{+}\Phi .
\end{eqnarray}%
In this formulation we may think of the real and complex Lagrangians, $%
\mathfrak{l}_{2}$ and $\mathcal{L}_{2}$, as being simply related by a kind
of Wick rotation in the field-configuration space 
\begin{equation}
\Phi ^{k}\rightarrow T\Phi ^{k}~,~~\Psi ^{k}\rightarrow T\Psi ^{k},~~\text{%
with }T:=\left( 
\begin{array}{cc}
1 & 0 \\ 
0 & -i%
\end{array}%
\right) .
\end{equation}

One may of course wonder about the negative sign in the kinetic term of $\mathfrak{l}_{2}$ and whether these lead to ghost fields. Thus we finish this subsection with a short discussion that establishes that these signs disappear when the Lagrangian is properly diagonalised. As this argument is the same in the global and local theory, we set the gauge fields to zero without loss of generality for this purpose and consider the corresponding action
\begin{eqnarray}
	S &=&\frac{1}{2}\int d^{4}x\left( \partial _{\mu }\Phi ^{\top}\mathcal{I}%
	\partial ^{\mu }\Phi -\Phi ^{\top}H_{\pm }\Phi \right) +S_{\text{int}}, \\
	&=&-\frac{1}{2}\int d^{4}x\Phi ^{\top}\mathcal{I}\left( \partial _{\mu
	}\partial ^{\mu }+M^{2}\right) \Phi +S_{\text{int}},  \label{S2}
\end{eqnarray}%
where $S_{\text{int}}$ contains all terms of higher order than $\Phi ^{2}$.
We also assumed that surface terms vanish at infinity when integrating by
parts and used $\mathcal{I}^{2}=\mathbb{I}$, $M^{2}=\mathcal{I}H_{\pm }$.
Next we diagonalise the squared mass matrix as $M^{2}=T^{-1}DT$ and consider next
only the integrant of the first term in (\ref{S2})
\begin{equation}
	\Phi ^{\top}\mathcal{I}\left( \partial _{\mu }\partial ^{\mu }+M^{2}\right)
	\Phi =\Phi ^{\top}\mathcal{I}\left( \partial _{\mu }\partial ^{\mu
	}+TDT^{-1}\right) \Phi =\Psi ^{\top}\left( \partial _{\mu
	}\partial ^{\mu }+D\right) \Psi ,
\end{equation}%
where we introduced the new field $\Psi :=$ $T^{\top}\mathcal{I}\Phi $ and used 
$T^{-1}=T^{\top}\mathcal{I}$.

The latter relation is derived as follows: We start by defining the right and left eigenvectors of $M^2$ as
\begin{equation}
	\mathcal{I}H_{\pm } v_i = \lambda_i v_i, \qquad \text{and} \qquad 
		(\mathcal{I}H_{\pm })^{\dagger} u_i = \lambda_i u_i,
\end{equation}%
respectively. Noting that $\mathcal{I}^\dagger=\mathcal{I} $ and $ H_{\pm }^\dagger =H_{\pm }$ the last relation implies that $\mathcal{I}H_{\pm } \mathcal{I} u_i =
\lambda_i \mathcal{I} u_i$. Therefore we can express the right eigenvectors in terms of the left eigenvectors as $v_i =\mathcal{I} u_i $. The matrix $T$ is made out of the column vectors of $v_i$, i.e. $T=(v_1,\ldots)$ so that $\mathcal{I} T= (u_1,\ldots)$. Since the left and right eigenvector form a biorthonormal basis, $v_i \cdot u_j =\delta_{ij}$, it follows that $T^{\top} \mathcal{I} T =\mathbb{I} $ and hence $T^{-1}=T^{\top}\mathcal{I}$. 

\subsubsection{The symmetry breaking vacuum}

The vacuum solutions $\Phi
_{0}^{k},\Psi _{0}^{k}$ by solving $\delta V=0$, 
	with $V$ denoting the potential
	\begin{equation}
		V= {\Phi}^{\top} H_+ \Phi + {\Psi}^{\top} H_- \Psi + \frac{g}{16} \left(  {\Phi}^{\top} E \Phi + {\Psi}^{\top} E \Psi    \right)^2 \label{potentials}
\end{equation}
amounts to solving
the two equations 
\begin{equation}
\left( H_{-}+\frac{g}{4}R^{2}E\right) \Psi _{0}^{k}=0~,~~\left( H_{+}+\frac{g%
}{4}R^{2}E\right) \Phi _{0}^{k}=0~,~~~\ ~~~k=1,2,  \label{vacuum equations}
\end{equation}%
with $R^{2}:=\left\vert \left( \phi _{1}^{0}\right) ^{1}\right\vert
^{2}+\left\vert \left( \phi _{1}^{0}\right) ^{2}\right\vert ^{2}=\frac{1}{2}%
\sum_{k=1}^{2}{\Phi _{0}^{k}}^{T}E\Phi _{0}^{k}+{\Psi _{0}^{k}}^{T}E\Psi
_{0}^{k}$=const. Hence in the real component field configuration space the
vacuum manifold is a $S^{3}$-sphere with radius $R$. Consequently, we may
consider the equations (\ref{vacuum equations}) as two eigenvalue equations.
Thus, besides the trivial $SU(2)$-invariant vacuum $\Phi _{0}^{k}=\Psi
_{0}^{k}=0$, $k=1,2$, we must have zero eigenvalues in both equations, which
is equivalent to requiring 
\begin{equation}
R^{2}=\frac{4}{gm_{2}^{2}}(\mu ^{4}+m_{1}^{2}m_{2}^{2}).
\end{equation}%
Since $R^{2}$ is positive, this equality imposes restrictions on the
parameters $g,\mu $ and the possible choices for $m_{1}\in \mathbb{R}$, $%
m_{2}\in i\mathbb{R}$ or $m_{1}\in i\mathbb{R}$, $m_{2}\in \mathbb{R}$. The
corresponding vectors that satisfy equation) (\ref{vacuum equations}), suitably normalized with regard to the standard inner product, are
\begin{equation}
\Psi _{0}^{2}=N_{\Psi }\left( 
\begin{array}{c}
m_{2}^{2} \\ 
\mu ^{2}%
\end{array}%
\right) ,~~\ \Phi _{0}^{2}=N_{\Phi }\left( 
\begin{array}{c}
-m_{2}^{2} \\ 
\mu ^{2}%
\end{array}%
\right) .  \label{phi}
\end{equation}%
Imposing now the constraint on $R^{2}$ as stated after equation (\ref{vacuum
equations}), a possible solution is $\Phi _{0}^{1}=\Psi _{0}^{1}=\Phi
_{0}^{2}=0$ and\ $\Psi _{0}^{2}$ as defined in (\ref{phi}) with
normalization constant $N_{\Psi }=\pm \sqrt{2}R/m_{2}^{2}$. Hence we recover
the symmetry breaking vacuum used in \cite{fring2020goldstone}.

\subsubsection{The Higgs mechanism}

Let us now demonstrate how the gauge vector boson acquires a finite mass and
how at the same time the emergence of a Goldstone boson is prevented by the
Higgs mechanism \cite%
{englert1964broken,higgs1964brokena,higgs1964brokenb,guralnik1964global} in
the $\mathcal{CPT}$-symmetric regime, at the exceptional points and even in
the spontaneously broken $\mathcal{CPT}$-symmetric regime. The mechanism
breaks down at the two types of zero exceptional points.

Expanding the potential $V$ in (\ref{potentials}) 
around the vacuum specified at the end of subsection
2.1.1 leads to 
\begin{eqnarray}
 V(\Phi_0 + \Phi,\Psi_0 + \Psi)	&=&  V(\Phi_0 ,\Psi_0 ) + 
\frac{1}{2} \Phi^i \left. \frac{\partial^2 V(\Phi_0,\Psi_0)}{\partial \Phi^i \partial \Phi^j} \right| \Phi^j
+ 
\frac{1}{2} \Psi^i \left. \frac{\partial^2 V(\Phi_0,\Psi_0)}{\partial \Psi^i \partial \Psi^j} \right| \Psi^j  \qquad\\ 
&&  +  \Phi^i \left. \frac{\partial^2 V(\Phi_0,\Psi_0)}{\partial \Phi^i \partial \Psi^j} \right| \Psi^j + \dots \notag \\
	 &=&\frac{1}{2}\sum_{i=1}^{2}-{\Phi ^{i}}^{\top}\left( H_{+}+\frac{g}{4}%
R^{2}E\right) \Phi ^{i}-{\Psi ^{1}}^{\top}\left( H_{-}+\frac{g}{4}R^{2}E\right)
\Psi ^{1}  \label{V} \\
&&-{\Psi ^{2}}^{\top}\left[ H_{-}+\frac{g}{4}R^{2}E+-\frac{g}{2}(E\Psi
_{0}^{2})^{2}E\right] \Psi ^{2}+\dots  \notag
\end{eqnarray}%
As expected, multiplying the Hessians in (\ref{V}) by $\mathcal{I}$ gives
back the squared mass matrix we found in \cite{fring2020goldstone}. The
kinetic term is almost unchanged except for the term involving $\Psi ^{2}$ 
\begin{eqnarray}
T &=&\frac{1}{2}\left[ \partial _{\mu }\Phi +e\mathcal{I}A_{\mu }\Psi \right]
^{\dagger }\mathcal{I}\left[ \partial ^{\mu }\Phi +e\mathcal{I}A^{\mu }\Psi %
\right] +\func{Re}\left\{ \left( \partial _{\mu }\Phi +e\mathcal{I}A_{\mu
}\Psi \right) ^{\dagger }\mathcal{I}\left( e\mathcal{I}A^{\mu }\Psi
_{0}\right) \right\} \label{kinetisch} \\
&&-\func{Im}\left\{ \left( \partial _{\mu }\Phi +e\mathcal{I}A_{\mu }\Psi +e%
\mathcal{I}A_{\mu }\Psi _{0}\right) ^{\dagger }\left( \partial ^{\mu }\Psi -e%
\mathcal{I}A^{\mu }\Phi \right) \right\} +\frac{1}{2}e^{2}{\left( A_{\mu
}\Psi _{0}\right) }^{\dagger }\mathcal{I}\left( A^{\mu }\Psi _{0}\right) 
\notag
\end{eqnarray}%
The last term corresponds to the mass term of the gauge vector boson that we
evaluate to 
\begin{eqnarray}
\frac{1}{2}e^{2}{\left( A_{\mu }\Psi _{0}\right) ^{\ast }}\mathcal{I}\left(
A^{\mu }\Psi _{0}\right) &=&\frac{1}{2}e^{2}{\left( A_{\mu }\Psi _{0}\right)
^{\ast }}_{\alpha }^{k}\mathcal{I}_{\alpha \beta }{\left( A^{\mu }\Psi
_{0}\right) }_{\beta }^{k} \\
&=&\frac{1}{2}e^{2}\left( A_{\mu }^{\dagger }A^{\mu }\right) ^{kj}\left( {%
\Psi _{0}}\right) _{\alpha }^{k}\mathcal{I}_{\alpha \beta }\left( {\Psi _{0}}%
\right) _{\beta }^{j}  \notag \\
&=&\frac{1}{2}e^{2}\left( A_{\mu }^{\dagger }A^{\mu }\right) ^{22}\left( {%
\Psi _{0}}\right) _{\alpha }^{2}\mathcal{I}_{\alpha \beta }\left( {\Psi _{0}}%
\right) _{\beta }^{2}  \notag \\
&=&\frac{1}{2}e^{2}A_{\mu }^{a}A^{b\mu }(\tau ^{a\dagger }\tau ^{b})^{22}%
\frac{2R^{2}}{m_{2}^{4}}\left( m_{2}^{4}-\mu ^{4}\right)  \notag \\
&=&\frac{1}{2}m_{g}^{2}A_{\mu }^{a}A^{a\mu },
\end{eqnarray}%
where we used the standard relation $\tau ^{a\dagger }\tau ^{b}=\tau
^{a}\tau ^{b}=\delta _{ab}\mathbb{I}+i\varepsilon _{abc}\tau ^{c}$.
Therefore we read off the mass of each of the three components of the gauge
vector boson as 
\begin{equation}
m_{g}:=\frac{\sqrt{2}eR}{m_{2}^{2}}\sqrt{m_{2}^{4}-\mu ^{4}}.  \label{gvb}
\end{equation}%
In \cite{fring2020goldstone} we identified the physical regions in the
parameter space in which the squared mass matrix has non-negative
eigenvalues and in which the Goldstone bosons can be identified. Let us now
compare those regions with the values for which the gauge vector boson
becomes massive. We immediately see from the expression in (\ref{gvb}) that
the gauge vector boson remains massless when $\mu ^{4}=m_{2}^{4}$ or when $%
R=0$, i.e. $\mu ^{4}=-m_{1}^{2}m_{2}^{2}$. The two sets of values correspond
precisely to the two types of zero exceptional points, type I and II,
respectively, at which the squared mass matrix develops zero eigenvalues $\lambda =0$.
These points are distinct from standard exceptional points where two
eigenvalues coalesce and become complex thereafter, here at $\lambda = \frac{\mu^4}{m_2^2}-m_2^2$. See the appendix for a more detailed explanation about the distinction between these types of exceptional points.

Thus the two aspects of
the Higgs-mechanism, i.e. giving mass to the gauge vector boson and at the
same time preventing the existence of the Goldstone bosons, remain to go
hand in hand. In the $\mathcal{CPT}$-symmetric regime the mechanism applies,
but at the zero exceptional points the Higgs-mechanism breaks down as the
Goldstone bosons are not identifiable\ \cite{fring2020goldstone} and at the
same time the gauge vector boson remains massless. In contrast, at the
exceptional point the Goldstone bosons are identifiable\ \cite%
{fring2020goldstone}, although in a different manner, and the gauge vector
bosons become massive.

Let us see this in detail by following \cite{fring2020goldstone} and
replacing $m_{i}^{2}\rightarrow c_{i}m_{i}^{2}$, with $c_{i}=\pm 1$ to
account for all possibilities in signs. We found that physical regions only
exist for the two cases $c_{1}=-$ $c_{2}=1$ and $c_{1}=-$ $c_{2}=-1$. For
the two cases we may then write%
\begin{equation}
\frac{m_{g}^{2}}{m_{1}^{2}}=c_{2}\frac{4e^{2}}{g}\frac{m_{1}^{6}}{m_{2}^{6}}%
\left( \frac{m_{2}^{4}}{m_{1}^{4}}-\frac{\mu ^{4}}{m_{1}^{4}}\right) \left( 
\frac{\mu ^{4}}{m_{1}^{4}}-\frac{m_{2}^{2}}{m_{1}^{2}}\right) ,  \label{mg}
\end{equation}%
noting that $m_{g}^{2}/m_{1}^{2}$ only depends on the two parameters $%
m_{2}^{2}/m_{1}^{2}$ and $m_{2}^{4}/m_{1}^{4}$ similarly as the
eigenspectrum of the squared mass matrix \cite{alex2019,fring2020goldstone}.
We require the right hand side of (\ref{mg}) to be positive as depicted in
depict in figure \ref{Fig1}.

\FIGURE{ \epsfig{file=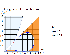, height=6.5cm} \epsfig{file=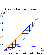, height=6.5cm}
\caption{Regions, for which the gauge vector boson is massive (blue with mesh) versus physical regions (orange) in which the would be Goldstone boson can be identified, bounded by exceptional and zero exceptional points as function of $(\mu^{4}/m_{1}^{4},m_{2}^{2}/m_{1}^{2})$ 
for the theory expanded around the SU(2)-symmetry breaking vacuum. Left panel for $c_1=-c_2=1$ and right panel for $c_1=-c_2=-1$. The coupling constant $g$ must be positive.}
       \label{Fig1}}

We observe in figure \ref{Fig1} that while the region in which the Goldstone
boson can be identified is bounded by exceptional as well as zero
exceptional points, the exceptional points lie well inside the region for
which the gauge vector boson is massive, i.e. they acquire a mass in the $%
\mathcal{CPT}$-symmetric regime as well as in the spontaneously broken $%
\mathcal{CPT}$-symmetric regime. In the $\mathcal{CPT}$-symmetric regime
this agrees well with the findings that at these points the
\textquotedblleft would be Goldstone boson\textquotedblright\ is prevented
from existing as a massless particle. We may think of the sign change in
front of the mass terms, $c_{i}\rightarrow -c_{i}$, that relates the left to
the right panel as a phase transition \cite{landau37}.

Let us now demonstrate this behaviour in detail and expand for this purpose
the Lagrangian around the symmetry broken vacuum up to second order in the
fields 
\begin{eqnarray}
\mathfrak{l}_{2} &=&\sum_{k=1}^{2}\frac{1}{2}\partial _{\mu }{\Phi ^{k}}^{T}%
\mathcal{I}\partial ^{\mu }\Phi ^{k}+\frac{1}{2}\partial _{\mu }{\Psi ^{k}}%
^{T}\mathcal{I}\partial ^{\mu }\Psi ^{k}-\frac{1}{2}{\Phi ^{k}}^{T}\left(
H_{+}+\frac{g}{4}R^{2}E\right) \Phi ^{k}  \label{second} \\
&&-\frac{1}{2}{\Psi ^{1}}^{T}\left( H_{-}+\frac{g}{4}R^{2}E\right) \Psi ^{1}-%
\frac{1}{2}{\Psi ^{2}}^{T}\left( H_{+}+\frac{g}{4}R^{2}E+\frac{g}{2}(E\Psi
_{(0)}^{2})^{2}E\right) \Psi ^{2}  \notag \\
&&+e\func{Re}\left[ \partial _{\mu }{\Phi }^{\dagger }(A^{\mu }\Psi _{0})%
\right] +e\func{Im}\left[ \left( \mathcal{I}A_{\mu }\Psi _{0}\right)
^{\dagger }\partial ^{\mu }\Psi \right] +\frac{1}{2}m_{g}^{2}A_{\mu
}^{a}A^{a\mu }+\dots  \notag
\end{eqnarray}%
We recall now from \cite{fring2020goldstone} that the first two lines of the
Lagrangian $\mathfrak{l}_{2}$ can be diagonalized and the Goldstone bosons
can be identified in terms of the field content of the model. Furthermore,
the Goldstone modes are eigenvectors in the null space of squared mass matrices 
\begin{equation}
M_{\pm }^{2}:=\mathcal{I}\left( H_{\pm }+\frac{g}{4}R^{2}E\right) ~,
\end{equation}%
computed above as $\Psi _{0}^{2}$ and $\mathcal{I}\Psi _{0}^{2}$, so that
the Goldstone modes are proportional to these two vectors. The explicit
forms of the Goldstone fields were found in \cite{fring2019pseudo} (equation (3.40) therein),
denoted as $\psi _{5}^{\text{Gb}}$, $\psi _{3}^{\text{Gb}}$and $\psi _{1}^{%
\text{Gb}}$, therein. We express them here as 
\begin{equation}
G^{1}:=\frac{e}{m_{g}}\left( {\Psi _{0}^{2}}\right) ^{T}\Phi ^{1}~,~~G^{3}:=%
\frac{e}{m_{g}}\left( {\Psi _{0}^{2}}\right) ^{T}\Phi ^{2}~,~~G^{2}:=-\frac{e%
}{m_{g}}\left( {\Psi _{0}^{2}}\right) ^{T}\mathcal{I}\Psi ^{1},
\end{equation}%
respectively. As expected for the Higgs mechanism the number of
\textquotedblleft would be Goldstone bosons\textquotedblright\ equals the
amount of massive vector gauge bosons. The fact that the Goldstone modes are
inverse proportional to the mass of the gauge bosons explains that they can
not be identified for massless gauge bosons. Keeping now only the Goldstone
kinetic term from the first two lines of the Lagrangian $\mathfrak{l}_{2}$
and the one involving the gauge fields in equation (\ref{second}), we obtain 
\begin{equation}
\mathfrak{l}_{2}=\sum_{a=1}^{3}\frac{1}{2}\partial _{\mu }G^{a}\partial
^{\mu }G^{a}+e\func{Re}\left[ \partial _{\mu }{\Phi }^{\dagger }(A^{\mu
}\Psi _{0})\right] +e\func{Im}\left[ \left( \mathcal{I}A_{\mu }\Psi
_{0}\right) ^{\dagger }\partial ^{\mu }\Psi \right] +\frac{1}{2}%
m_{g}^{2}A_{\mu }^{a}A^{a\mu }+\dots  \label{GA}
\end{equation}%
Using the explicit representations of the Pauli matrices, the real and
imaginary parts are determined as 
\begin{eqnarray}
\func{Re}\left[ \partial _{\mu }\Phi ^{T}A^{\mu }\Psi _{0}\right] &=&A_{\mu
}^{a}\func{Re}\left[ \partial _{\mu }\Phi ^{T}\tau ^{a}\Psi _{0}\right]
=A_{\mu }^{1}\partial _{\mu }\Phi ^{T}\tau ^{1}\Psi _{0}+A_{\mu
}^{3}\partial _{\mu }\Phi ^{T}\tau ^{3}\Psi _{0} \\
&=&A_{\mu }^{1}\partial _{\mu }\left( {\Phi ^{1}}\right) ^{T}{\Psi _{0}^{2}}%
-A_{\mu }^{3}\partial _{\mu }\left( {\Phi ^{2}}\right) ^{T}{\Psi _{0}^{2}} 
\notag \\
&=&A_{\mu }^{1}\frac{m_{g}}{e}\partial ^{\mu }G^{1}-\frac{m_{g}}{e}A_{\mu
}^{3}\partial ^{\mu }G^{3}  \notag \\
\func{Im}\left[ \left( \mathcal{I}A_{\mu }\Psi _{0}\right) ^{\dagger
}\partial ^{\mu }\Psi \right] &=&A_{\mu }^{a}\func{Im}\left[ \Psi
_{0}^{T}\tau ^{a}\mathcal{I}\partial ^{\mu }\Psi \right] =-i\left( A_{\mu
}^{2}\Psi _{0}^{T}\tau ^{2}\mathcal{I}\partial ^{\mu }\Psi \right) \\
&=&A_{\mu }^{2}\left( {\Psi _{0}^{2}}\right) ^{T}\mathcal{I}\partial ^{\mu
}\Psi ^{1}=-A_{\mu }^{2}\frac{m_{g}}{e}\partial ^{\mu }G^{2}.  \notag
\end{eqnarray}%
Finally the Lagrangian in (\ref{GA}) can be simplified to%
\begin{eqnarray}
\mathfrak{l}_{2} &=&\sum_{a=1}^{3}\frac{1}{2}\partial _{\mu }G^{a}\partial
^{\mu }G^{a}-m_{g}A_{\mu }^{1}\partial ^{\mu }G^{1}+m_{g}A_{\mu
}^{2}\partial ^{\mu }G^{2}-m_{g}A_{\mu }^{3}\partial ^{\mu }G^{3}+\frac{1}{2}%
m_{g}^{2}A_{\mu }^{a}A^{a\mu }+\dots \\
&=&\frac{1}{2}m_{g}^{2}\left( A_{\mu }^{1}-\frac{1}{m_{g}}\partial _{\mu
}G^{1}\right) ^{2}+\frac{1}{2}m_{g}^{2}\left( A_{\mu }^{2}+\frac{1}{m_{g}}%
\partial _{\mu }G^{2}\right) ^{2}+\frac{1}{2}m_{g}^{2}\left( A_{\mu }^{3}+%
\frac{1}{m_{g}}\partial _{\mu }G^{3}\right) ^{2}+\dots  \notag \\
&=&\frac{1}{2}\sum_{a=1}^{3}m_{g}^{2}B_{\mu }^{a}B^{a\mu }+\dots ,  \notag
\end{eqnarray}%
where we defined the new vector gauge particle with component fields $B_{\mu
}^{a}:=A_{\mu }^{a}-\frac{1}{m_{g}}\partial _{\mu }G^{a}$. We may also
replace $A_{\mu }^{a}$ by $B_{\mu }^{a}$ in the field strength $F_{\mu \nu }$
so that $A_{\mu }$ can be eliminated entirely from the Lagrangian. We see
that the Higgs-mechanism applies as long as $m_{g}\not=0$. However, at the
zero exceptional points, not only the gauge boson mass vanishes, but the
Higgs mechanism no longer applies, in the sense that we can not remove the
degrees of freedom of Goldstone bosons.

We summarize the behaviour we found in the different types of regimes in the
following table

\begin{center}
\begin{tabular}{|l|l|l|l|l|l|}
\hline
& \multicolumn{1}{|l|}{$\mathcal{CPT}$} & \multicolumn{1}{|l|}{sp. broken $%
\mathcal{CPT}$} & EP & zero EP I & zero EP II \\ \hline\hline
gauge bosons & \multicolumn{1}{|c|}{massive} & \multicolumn{1}{|c|}{massive}
& \multicolumn{1}{|c|}{massive} & \multicolumn{1}{|c|}{massless} & 
\multicolumn{1}{|c|}{massless} \\ \hline
Goldstone bosons & \multicolumn{1}{|c|}{$\exists $} & \multicolumn{1}{|c|}{$%
\exists $} & \multicolumn{1}{|c|}{$\exists $} & \multicolumn{1}{|c|}{$%
\nexists $} & \multicolumn{1}{|c|}{$\nexists $} \\ \hline
\end{tabular}
\end{center}

Thus we encounter three different types of behaviour: In the $\mathcal{CPT}$%
-symmetric regime, at the standard exceptional points as well as in the
spontaneously broken $\mathcal{CPT}$-symmetric regime the Higgs mechanism
applies in the usual way. However, in the latter regime other particles in
the theory become non-physical. At the zero exceptional points the vector
gauge bosons remain massless and no Goldstone bosons can be identified in
the global theory.

\subsubsection{From $SU(2)$ to $SU(N)$}

We will now follow the same line of reasoning as in the previous subsection
and generalize our model from possessing a $SU(2)$-symmetry to one with a $%
SU(N)$-symmetry. For this purpose we simply replace the Pauli matrices in
all our expressions by the traceless and skew-Hermitian $N\times N$-matrices
corresponding to the $SU(N)$-generators $T^{a}$ with $a=1,\ldots ,(N^{2}-1)$%
. The vacua are still determined by the solutions of the eigenvalue problem (%
\ref{vacuum equations}) with zero eigenvalue condition 
\begin{equation}
R^{2}=\frac{1}{2}\sum_{i=1}^{N}{\Phi _{0}^{i}}^{T}E\Phi _{0}^{i}+{\Psi
_{0}^{i}}^{T}E\Psi _{0}^{i}=\text{constant}=\frac{4}{gm_{2}^{2}}(\mu
^{4}+m_{1}^{2}m_{2}^{2}).  \label{R2}
\end{equation}%
The zero eigenvalue condition implies that the vacuum manifold is a $%
S^{2N-1} $-sphere with radius $R$. This follows from the fact that $SU(N)$
acts on the $2N$ dimensional space spanned by $(\varphi _{1}^{0})^{i}$, $%
(\chi _{1}^{0})^{i},i=1,\ldots ,N$, with norm equal to $R^{2}$. On this
space $SU(N-1)$ simply permutes the fields amongst themselves, hence acting
as a stabilizer or isotropy subgroup. Thus the vacuum manifold corresponds to
the coset $SU(N)/SU(N-1)\cong S^{2N-1}$.

As we discussed in detail in \cite{fring2020goldstone}, we may utilize the
symmetry of the Lagrangian to transform the vacua into convenient forms
without changing the eigenvalue spectrum of the mass matrix. Thus using the
elements $T\in SU(N)/SU(N-1)\subset SU(N)$ we may transform the vacuum
into the form 
\begin{equation}
\Phi _{0}^{i}=0,~~~~\Psi _{0}^{i}=\frac{\sqrt{2}R}{\sqrt{N}m_{2}^{2}}\left( 
\begin{array}{c}
m_{2}^{2} \\ 
\mu ^{2}%
\end{array}%
\right) ,~~~\ \text{for }i=1,\ldots ,N,
\end{equation}%
satisfying the constraint (\ref{R2}). Let us now use this $SU(N)$-symmetry
breaking vacuum to calculate the mass of the gauge vector boson. Taking the
proper $SU(N)$-algebra rather than the physicist's version, as in the last
subsection for $SU(2)$, we also change $e\rightarrow ie$.  Dropping here the kinetic term reported in (\ref{kinetisch}) and considering only
the relevant term in the Lagrangian we obtain 
\begin{eqnarray}
\mathfrak{l}_{A} &:&=\frac{1}{2}e^{2}A_{\mu }^{a}A^{b\mu }\left( {T^{a}}%
^{\dagger }T^{b}\right) _{ij}\left( {\Psi _{0}}\right) _{\alpha }^{i}%
\mathcal{I}_{\alpha \beta }\left( {\Psi _{0}}\right) _{\beta }^{j} \\
&=&\frac{1}{N}e^{2}A_{\mu }^{a}A^{b\mu }R^{2}\left( 1-\frac{\mu ^{4}}{%
m_{2}^{4}}\right) \sum\nolimits_{i,j=1}^{N}\left( {T^{a}}T^{b}\right) _{ij}.
\notag
\end{eqnarray}%
We evaluate the last factor using the identity ${T^{a}}T^{b}=\frac{1}{2N}%
\delta _{ab}\mathbb{I}_{N}+\frac{1}{2}\sum_{c=1}^{N^{2}-1}\left(
f_{abc}+ig_{abc}\right) T^{c},$ where the $g_{abc}$ and $f_{abc}$ are
completely symmetric and anti-symmetric tensors, respectively. We note that $%
\sum\nolimits_{i,j=1}^{N}(T^{c})_{ij}=\limfunc{Tr}T^{c}=0$ due to the
skew-Hermitian nature of $T^{c}$ and\ $\sum\nolimits_{i,j=1}^{N}(\mathbb{I}%
_{N})_{ij}=\limfunc{Tr}\mathbb{I}_{N}=N$. Thus we can diagonalize $\mathfrak{%
l}_{A}$, computing%
\begin{equation}
\mathfrak{l}_{A}=\frac{R^{2}}{2N}e^{2}\left( 1-\frac{\mu ^{4}}{m_{2}^{4}}%
\right) A_{\mu }^{a}A^{a\mu }=\frac{1}{2}m_{g}^{2}A_{\mu }^{a}A^{a\mu },
\end{equation}%
from which we read off the masses $m_{g}^{(a)}$ of the $N^{2}-1$ gauge
vector bosons. We note that once again they vanish at the zero exceptional
points, but now for all $SU(N)$-models.

\subsection{A $SU(2)$-symmetric model in adjoint representation}

As we have demonstrated, the gauge vector boson becomes massive for the $%
SU(N)$-symmetric model in the $\mathcal{CPT}$-symmetric regime and at the
exceptional point when the fields are taken to be in the representation
space of the fundamental representation. On the other hand the
Higgs-mechanism breaks down at the zero exceptional points. Remarkably it
still applies when the $\mathcal{CPT}$-symmetry is broken, although in that
regime other particles acquire complex masses so that the region is
non-physical. Recall that the regions in which the $\mathcal{CPT}$-symmetry holds are identified in figure 1.

Let us now see whether we encounter a similar behaviour when
the fields are taken in adjoint representation. We consider here a slightly
different non-Hermitian $SU(2)$-invariant Lagrangian 
\begin{eqnarray}
\mathcal{L}_{2}^{\text{ad}} &=&\frac{1}{2}\limfunc{Tr}\left( D\phi
_{1}\right) ^{2}+\frac{1}{2}\limfunc{Tr}\left( D\phi _{2}\right) ^{2}-\frac{%
m_{1}^{2}}{2}\limfunc{Tr}(\phi _{1}^{2})+\frac{m_{2}^{2}}{2}\limfunc{Tr}%
(\phi _{2}^{2})-i\mu ^{2}\limfunc{Tr}(\phi _{1}\phi _{2})
\label{adj_rep_action} \\
&&-\frac{g}{4}\left[ \limfunc{Tr}(\phi _{1}^{2})\right] ^{2}-\frac{1}{4}%
\limfunc{Tr}\left( F^{2}\right) ,  \notag
\end{eqnarray}%
where as in equation (\ref{LN}) we take $g,\mu \in \mathbb{R}$, $m_{i}\in 
\mathbb{R}$ or $m_{i}\in i\mathbb{R}$, to be constants. The two complex
scalar fields are expressed as $\phi _{i}=\phi _{i}^{a}T^{a}$, $i=1,2$ and $%
a=1,2,3$, where the $T^{a}$ are the three $SU(2)$-generators in the adjoint
representation that, up to a factor of $2$, satisfy the same algebra as the
Pauli spin matrices, that is $[T^{a},T^{b}]=i\varepsilon _{abc}T^{c}$.
Hence, the adjoint representation is $\left( T^{a}\right)
_{bc}=-i\varepsilon _{abc}$, i.e. to be explicit 
\begin{equation}
T^{1}=\left( 
\begin{array}{ccc}
0 & 0 & 0 \\ 
0 & 0 & -i \\ 
0 & i & 0%
\end{array}%
\right) ,\quad T^{2}=\left( 
\begin{array}{ccc}
0 & 0 & i \\ 
0 & 0 & 0 \\ 
-i & 0 & 0%
\end{array}%
\right) ,\quad T^{3}=\left( 
\begin{array}{ccc}
0 & -i & 0 \\ 
i & 0 & 0 \\ 
0 & 0 & 0%
\end{array}%
\right) ,  \label{adT}
\end{equation}%
such that $\limfunc{Tr}(T^{a}T^{b})=2\delta ^{ab}$~and therefore $Tr(\phi
^{2})=2\sum_{a=1}^{3}\phi ^{a}\phi ^{a}$. The $SU(2)$-symmetry in the
adjoint representation for each generator $T^{a}$ is therefore%
\begin{equation}
\phi _{j}\rightarrow e^{i\alpha T^{a}}\phi _{j}e^{-i\alpha T^{a}}\approx
\phi _{j}-\alpha \varepsilon _{abc}\phi _{j}^{b}T^{c},  \label{s2ad}
\end{equation}%
so that the infinitesimal changes to the fields $\phi _{i}^{a}$ result to 
\begin{equation}
\delta \phi _{i}^{a}=-\alpha \varepsilon _{abc}\phi _{i}^{b}.  \label{infphi}
\end{equation}%
We will utilize this expression below.

In more a compact form the Lagrangian in (\ref{adj_rep_action}) can be
expressed equivalently as 
\begin{equation}
\mathcal{L}_{2}^{\text{ad}}=D_{\mu }\phi _{i}^{a}D^{\mu }\phi _{i}^{a}-\phi
_{i}^{a}M_{ij}^{2}\phi _{j}^{a}-g\left( \phi _{i}^{a}E_{ij}\phi
_{j}^{a}\right) ^{2}-\frac{1}{4}F_{\mu \nu }^{a}\left( F^{\mu \nu }\right)
^{a},  \label{adj_rep_action_2}
\end{equation}%
where repeated indices are summed over the appropriate index sets $i,j,\mu
,\nu \in \{1,2\}$ and $a,b\in \{1,2,3\}$. The matrix $M^{2}$ is defined as 
\begin{equation}
M^{2}=\left( 
\begin{array}{cc}
m_{1}^{2} & i\mu ^{2} \\ 
i\mu ^{2} & -m_{2}^{2}%
\end{array}%
\right) ~,~
\end{equation}%
and $E$ as in (\ref{HIE}). The covariant derivative in the adjoint
representation acting on a complex field takes on the form 
\begin{equation}
(D_{\mu }\phi _{i})^{a}:=\partial _{\mu }\phi _{i}^{a}+e\varepsilon
_{abc}A_{\mu }^{b}\phi _{i}^{c}
\end{equation}%
Pursuing here a pseudo-Hermitian approach we perform a similarity
transformation on the Lagrangian in (\ref{adj_rep_action_2}) with Dyson map 
\begin{equation}
\eta =\prod_{a=1}^{3}e^{\frac{\pi }{2}\int d^{3}x\Pi _{2}^{a}\phi _{2}^{a}},
\end{equation}%
that maps the complex Lagrangian $\mathcal{L}_{2}^{\text{ad}}$ to a real
Lagrangian 
\begin{equation}
\mathfrak{l}_{2}^{\text{ad}}=(D_{\mu }\phi _{i})^{a}\mathcal{I}_{ij}(D^{\mu
}\phi _{j})^{a}-\phi _{i}^{a}H_{ij}\phi _{j}^{a}-g\left( \phi
_{i}^{a}E_{ij}\phi _{j}^{a}\right) ^{2},  \label{adj_rep_action_real}
\end{equation}%
where the matrix $H$ is defined as 
\begin{equation}
H:=\left( 
\begin{array}{cc}
m_{1}^{2} & \mu ^{2} \\ 
\mu ^{2} & m_{2}^{2}%
\end{array}%
\right) ~,~
\end{equation}%
and $\mathcal{I}$ as in (\ref{HIE}).

\subsubsection{The $SU(2)$-symmetry preserving and breaking vacua}

To find the different types of vacua $\phi ^{0}$, we need to solve again $%
\delta V=0$. The corresponding functional variation of the Lagrangian in (%
\ref{adj_rep_action_real}) leads to the three sets of equations%
\begin{equation}
\left( H+2gR^{2}E\right) \left( \phi ^{0}\right) ^{a}=0,~~\ \ \ a=1,2,3,
\label{EVE}
\end{equation}%
with $R^{2}:=\left( \phi _{i}^{0}\right) ^{a}E_{ij}\left( \phi
_{j}^{0}\right) ^{a}$. Next to the trivial $SU(2)$-symmetry preserving
solution $\left( \phi ^{0}\right) ^{a}=0$, a $SU(2)$-symmetry breaking
solution is obtained by requiring $\left( \phi ^{0}\right) ^{a}$ to become a vector in the null space of the matrix $H+2gR^{2}E$, which is the case when \qquad\ 
\begin{equation}
\left( \phi ^{0}\right) ^{a}=\frac{N_{a}}{m_{2}^{2}}~\left( 
\begin{array}{c}
m_{2}^{2} \\ 
-\mu ^{2}%
\end{array}%
\right) ,~~\text{and\quad }R^{2}=\frac{\mu ^{4}-m_{1}^{2}m_{2}^{2}}{%
2gm_{2}^{2}},  \label{null}
\end{equation}%
where the $N_{a}$~are normalization constants. Given the 
solution in (\ref{null}), the relation for $R^{2}$ imposes the additional
constraint $R^{2}=N_{1}^{2}+N_{2}^{2}+N_{3}^{2}$ on these constants.
Expressing the Lie algebra valued vacuum field $\phi _{i}^{0}=\left( \phi
_{i}^{0}\right) ^{a}T^{a}$ in the matrix form of the adjoint representation (%
\ref{adT}) we obtain%
\begin{equation}
\phi _{1}^{0}=i\left( 
\begin{array}{ccc}
0 & -N_{3} & N_{2} \\ 
N_{3} & 0 & -N_{1} \\ 
-N_{2} & N_{1} & 0%
\end{array}%
\right) ~,~~\text{and\quad }\phi _{2}^{0}=-\frac{\mu ^{2}}{m_{2}^{2}}\phi
_{1}^{0}.
\end{equation}%
We can now apply the $SU(2)$-symmetry to the vacuum state in the form 
\begin{equation}
\phi ^{\text{vac}}=\left[ \left( \phi _{1}^{0}\right) ^{1},\left( \phi
_{2}^{0}\right) ^{1},\left( \phi _{1}^{0}\right) ^{2},\left( \phi
_{2}^{0}\right) ^{2},\left( \phi _{1}^{0}\right) ^{3},\left( \phi
_{2}^{0}\right) ^{3}\right] ,  \label{sbvac}
\end{equation}%
so that the infinitesimal changes $\delta \phi _{i}(\phi ^{\text{vac}})$
with (\ref{infphi}) and (\ref{null}) yield the following states for each
generator%
\begin{eqnarray}
v_{1}^{0} &=&\frac{\alpha _{1}}{m_{2}^{2}}\left(
0,0,N_{3}m_{2}^{2},-N_{3}\mu ^{2},-N_{2}m_{2}^{2},N_{2}\mu ^{2}\right) ,
\label{v1} \\
v_{2}^{0} &=&\frac{\alpha _{2}}{m_{2}^{2}}\left( -N_{3}m_{2}^{2},N_{3}\mu
^{2},0,0,N_{1}m_{2}^{2},-N_{1}\mu ^{2}\right) ,  \label{v2} \\
v_{3}^{0} &=&\frac{\alpha _{3}}{m_{2}^{2}}\left( N_{2}m_{2}^{2},-N_{2}\mu
^{2},-N_{1}m_{2}^{2},N_{1}\mu ^{2},0,0\right) ,  \label{v3}
\end{eqnarray}%
as solutions for $\phi ^{\text{vac}}$. Evidently, these states are linearly dependent as%
\begin{equation}
\sum\nolimits_{i=1}^{3}\frac{N_{i}v_{i}^{0}}{\alpha _{i}}=0.
\end{equation}%
According to Goldstone's theorem the states $v_{i}^{0}$ should be eigenvectors of the squared mass matrix with eigenvalue 0. As only two of them are linearly
independent we expect to find two massless Goldstone bosons, which in our
gauged model correspond to \textquotedblleft would be Goldstone
bosons\textquotedblright . Hence the $SU(2)$-symmetry has been broken down
to a $U(1)$-symmetry, so that the group theoretical argument predicts two
Goldstone bosons equal to the dimension of the coset $SU(2)/U(1)$.

\subsubsection{The squared mass matrix}

Expanding the Lagrangian in equation (\ref{adj_rep_action_2}) about the
vacuum solution gives 
\begin{equation}
\mathfrak{l}_{2}^{\text{ad}}=(D_{\mu }\phi _{i})^{a}\mathcal{I}_{ij}(D_{\mu
}\phi _{j})^{a}-\frac{1}{2}\phi _{i}^{a}H_{ij}^{(a)}\phi _{j}^{a}+2(D_{\mu
}\phi _{i}^{0})^{a}\mathcal{I}_{ij}(D^{\mu }\phi _{j})^{a}+(D_{\mu }\phi
_{i}^{0})^{a}\mathcal{I}_{ij}(D_{\mu }\phi _{j}^{0})^{a}+\mathcal{O}(\phi
^{3}),  \label{expanded_real_action}
\end{equation}%
where the last two terms originate from expanding the covariant kinetic
term. The Hessian matrix is then computed by differentiating (\ref{EVE})
once more%
\begin{equation}
\hat{H}_{ij}^{ab}:=\frac{\partial ^{2}\mathcal{V}}{\partial \phi
_{i}^{a}\partial \phi _{j}^{b}}=2H_{ij}+4gR^{2}E_{ij}\delta ^{ab}+8g\left(
E\phi ^{a}\right) _{i}\left( E\phi ^{b}\right) _{j},
\end{equation}%
from which we obtain the non-Hermitian squared mass matrix as%
\begin{eqnarray}
M^{2} &=&\left. \frac{1}{2}\mathcal{I}\hat{H}\right\vert _{\phi ^{\text{vac}%
}} \\
&=&\left( 
\begin{array}{cccccc}
m_{1}^{2}+2gR^{2}+4gN_{1}^{2} & \mu ^{2} & 4gN_{1}N_{2} & 0 & 4gN_{1}N_{3} & 
0 \\ 
-\mu ^{2} & -m_{2}^{2} & 0 & 0 & 0 & 0 \\ 
4gN_{1}N_{2} & 0 & m_{1}^{2}+2gR^{2}+4gN_{2}^{2} & \mu ^{2} & 4gN_{2}N_{3} & 
0 \\ 
0 & 0 & -\mu ^{2} & m_{2}^{2} & 0 & 0 \\ 
4gN_{1}N_{3} & 0 & 4gN_{2}N_{3} & 0 & m_{1}^{2}+2gR^{2}+4gN_{3}^{2} & \mu
^{2} \\ 
0 & 0 & 0 & 0 & -\mu ^{2} & -m_{2}^{2}%
\end{array}%
\right) .  \notag
\end{eqnarray}%
The entries in the rows and columns of $M^{2}$ are labeled as $(\phi
_{1}^{1},\phi _{2}^{1},\phi _{1}^{2},\phi _{2}^{2},\phi _{1}^{3},\phi
_{2}^{3})=:\Psi $. The six eigenvalues $\lambda $ of $M^{2}$ are then
computed to 
\begin{equation}
\lambda _{1,2}=0;\qquad \lambda _{3,4}=\frac{\mu ^{4}-m_{2}^{4}}{m_{2}^{2}}%
,\qquad \lambda _{\pm }=\kappa \pm \sqrt{2(\mu
^{4}-m_{1}^{2}m_{2}^{2})+\kappa ^{2}},  \label{EVl}
\end{equation}%
with $\kappa :=3\mu ^{4}/2m_{2}^{2}-m_{2}^{2}/2-m_{1}^{2}$. We can now
verify that the three vectors $v_{i}^{0}$ in (\ref{v1})-(\ref{v3}),
corresponding to the infinitesimal changes of the vacuum (\ref{null}) under
the action of the $SU(2)$-symmetry, are indeed vectors in the null space of  $M^{2}$. Due
to their linear dependence we may choose two of them to be associated with
the two massless \textquotedblleft would be Goldstone
bosons\textquotedblright .

We note that there are zero exceptional points at $\mu ^{4}=m_{2}^{4}$ when $%
\lambda _{3,4}=0$, and at $\mu ^{4}=m_{1}^{2}m_{2}^{2}$ when either $\lambda
_{-}=0$ or $\lambda _{+}=0$. The standard exceptional point for which the
two eigenvalues $\lambda _{-}$ and $\lambda _{+}$ coalesce occurs when $%
m_{1}^{2}=3\mu ^{4}/2m_{2}^{2}+m_{2}^{2}/2\pm \mu ^{2}$. The Jordan normal form for the mass squared matrix becomes
\begin{equation}
	\limfunc{diag}D_{e}=(0,\lambda _{e}^{b},0,\lambda _{e}^{b},0,\lambda
	_{e}^{b},\Lambda ),\text{~~~}\lambda _{e}^{b}=\frac{\mu ^{4}}{m_{2}^{2}}%
	-m_{2}^{2},~~~\Lambda =\left( 
	\begin{array}{ll}
		\pm \mu ^{2}-m_{2}^{2} & \pm (\alpha -\beta )\mu ^{2} \\ 
		0 & \pm \mu ^{2}-m_{2}^{2}%
	\end{array}%
	\right) ,
\end{equation}
for some arbitrary constants $\alpha$ and $\beta$.

 We notice that the
eigenvalues in (\ref{EVl}) do not depend on the choice of the three
normalization constants $N_{a}$, since all of these vacua are equivalent as
they are related by $SU(2)$-symmetry transformations. The physical regions
of the model are determined by the requirement that the eigenvalues are real
and positive. Taking now account of the possibility that $m_{i}\in \mathbb{R}
$ or $m_{i}\in i\mathbb{R}$, by allowing for different signs in front of the 
$m_{i}^{2}$ terms in setting $m_{i}^{2}\rightarrow c_{i}m_{i}^{2}$, we find
that the model does not possess any physical region when $c_{1}=c_{2}=\pm 1$
and physical regions when $c_{1}=-c_{2}=\pm 1$ as argued also in the
previous section.

\subsubsection{The would be Goldstone bosons}

Let us now identify the two massless Goldstone bosons $\psi _{1,2}^{\text{Gb}%
}$ in the different $\mathcal{PT}$ -regimes by the same procedure as
previously explained in \cite{fring2019pseudo,fring2020goldstone}, with the
difference that they will be made to vanish due to the presence of the gauge
bosons. In terms of the original scalar fields in the model we identify the
Goldstone bosons by evaluating%
\begin{equation}
\psi _{1,2}^{\text{Gb}}:=\sqrt{(\Psi ^{T}\hat{I}U)_{1,2}(U^{-1}\Psi )_{1,2}},
\label{Gold}
\end{equation}%
where the matrix $U$ diagonalizes the squared mass matrix by $U^{-1}M^{2}U=D$
with $\limfunc{diag}D=(\lambda _{1},\lambda _{2},\lambda _{3},\lambda
_{4},\lambda _{-},\lambda _{+})$ and $\limfunc{diag}\hat{I}=\{\mathcal{I},%
\mathcal{I},\mathcal{I}\}$. In the $\mathcal{PT}$ -symmetric regime the
similarity transformation $U$ is well defined by 
\begin{equation}
U:=(v_{1},v_{2},v_{3},v_{4},v_{-},v_{+}),  \label{FMU}
\end{equation}%
where the $v_{i}$ are the eigenvectors of $M^{2}$. Up to normalizations
constants for each eigenvector, we obtain in our example the concrete
expressions%
\begin{equation}
v_{i}=\left[ \left( m_{2}^{2}+\lambda _{i}\right) \tau _{i1},-\mu ^{2}\tau
_{i1},\left( m_{2}^{2}+\lambda _{i}\right) \tau _{i2},-\mu ^{2}\tau
_{i2},\left( m_{2}^{2}+\lambda _{i}\right) \tau _{i3},-\mu ^{2}\tau _{i3}%
\right] ,  \label{vec}
\end{equation}%
with $\tau _{12}=\tau _{23}=\tau _{32}=\tau _{43}=0$, $\tau _{33}=\tau
_{42}=\tau _{\pm 1}=-\tau _{13}=-\tau _{22}=N_{1}$, $\tau _{21}=\tau
_{41}=\tau _{\pm 2}=N_{2}$ and $\tau _{11}=\tau _{31}=\tau _{\pm 3}=N_{3}$.
Defining a $ \mathcal{PT}$-inner product as $\left\langle a|b\right\rangle _{ 
\mathcal{PT}}:=a\hat{I}b$ these vectors can be orthonormalized $%
\left\langle v_{i}|v_{j}\right\rangle _{ \mathcal{PT}}=\delta _{ij}$. Recall from the argument after (2.19) that indeed $v_i =\mathcal{I} u_i $, so that $%
\left\langle v_{i}|v_{j}\right\rangle _{ \mathcal{PT}}= v_{i} \mathcal{I}^2 u_j = \delta _{ij}$. Here we should stress that $M^2$ is only the analogue to a Hamiltonian and we are not attempting to construct an associated Hilbert space. 

For
convenience we take now $N_{1}=N_{2}=0$, $N_{3}=R$ and compute%
\begin{equation}
\psi _{1}^{\text{Gb}}:=\frac{m_{2}^{2}\phi _{1}^{3}+\mu ^{2}\phi _{2}^{3}}{%
\sqrt{m_{2}^{4}-\mu ^{4}}},\quad \text{and\quad }\psi _{2}^{\text{Gb}}:=%
\frac{m_{2}^{2}\phi _{1}^{2}+\mu ^{2}\phi _{2}^{2}}{\sqrt{m_{2}^{4}-\mu ^{4}}%
}.  \label{G12}
\end{equation}%
We note that $\det U=\lambda _{3}\lambda _{4}(\lambda _{-}-\lambda _{+})\mu
^{6}R^{4}$, indicating the breakdown of these expressions at the exceptional
points when $\lambda _{-}=\lambda _{+}$, the zero exceptional point when $%
\lambda _{3}=\lambda _{4}=0$ and at the trivial vacuum when $R=0$, as
previously observed in \cite{fring2019pseudo,fring2020goldstone}. However,
at the exceptional point we may still calculate the expressions for the
Goldstone boson when taking into account that in this case the two
eigenvectors $v_{-}$ and $v_{+}$ become identical. In order to obtain two
linearly independent eigenvectors when the squared mass matrix is converted
into its Jordan normal form we multiply two entries of the vector $v_{+}$ by
some arbitrary constants $\alpha \neq \beta $ as $(v_{+})_{1}\rightarrow
\alpha (v_{+})_{1}$ and $(v_{+})_{2}\rightarrow \beta (v_{+})_{2}$. With
this change the matrix $U$ becomes invertible as $\det U=\lambda _{3}\lambda
_{4}(\beta -\alpha )(m_{2}^{2}+\kappa )N_{1}^{2}\mu ^{6}R^{2}$. We may now
evaluate the expression in (\ref{Gold}) obtaining the same formulae for the
Goldstone bosons as in (\ref{G12}). At the zero exceptional point it is not
possible to identify the Goldstone in terms of the original fields in the
model.

\subsubsection{The mass of the vector gauge boson}

Finally we calculate the mass of the gauge vector bosons by expanding the
minimal coupling term in equation (\ref{adj_rep_action_real}) around the
symmetry breaking vacuum (\ref{sbvac}) 
\begin{eqnarray}
\left[ D_{\mu }(\phi +\phi ^{0})\right] ^{T}\mathcal{I}\left[ D^{\mu }(\phi
+\phi ^{0})\right] &=&(D_{\mu }\phi ^{0})^{T}\mathcal{I}(D^{\mu }\phi
^{0})+\dots  \label{gauge} \\
&=&e^{2}\left[ \varepsilon _{abc}A_{\mu }^{b}\left( \phi _{i}^{0}\right) ^{c}%
\right] \mathcal{I}_{ij}\left( \varepsilon _{ade}{A^{d}}^{\mu }\left( \phi
_{j}^{0}\right) ^{e}\right) +\dots  \notag \\
&=&e^{2}\left( A_{\mu }^{a}A^{a\mu }\left( \phi _{i}^{0}\right) ^{b}\mathcal{%
I}_{ij}\left( \phi _{j}^{0}\right) ^{b}-A_{\mu }^{a}A^{b\mu }\left( \phi
_{i}^{0}\right) ^{b}\mathcal{I}_{ij}\left( \phi _{j}^{0}\right) ^{a}\right)
+\dots ,  \notag
\end{eqnarray}%
where we used the standard identity $\varepsilon _{abc}\varepsilon
_{ade}=\delta _{bd}\delta _{ce}-\delta _{be}\delta _{cd}$. A convenient
choice for the normalization constants $N_{i}$ that is compatible with (\ref%
{null}) and diagonalizes (\ref{gauge}) is to set two constants to zero and
the remaining one to $R$. For instance, taking $N_{1}=N_{2}=0$, $N_{3}=R$
the only nonvanishing terms in (\ref{gauge}) are 
\begin{eqnarray}
&=&e^{2}\left( A_{\mu }^{1}A^{1\mu }+A_{\mu }^{2}A^{2\mu }\right) \left(
\phi _{i}^{0}\right) ^{3}\mathcal{I}_{ij}\left( \phi _{j}^{0}\right) ^{3}, \\
&=&e^{2}R^{2}\left( 1-\frac{\mu ^{4}}{m_{2}^{4}}\right) \left( A_{\mu
}^{1}A^{1\mu }+A_{\mu }^{2}A^{2\mu }\right) .
\end{eqnarray}%
Thus for $\mu ^{4}\neq m_{2}^{4}$ and $R\neq 0$ we obtain two massive vector
gauge bosons $m_{g}^{(1)}$ and $m_{g}^{(2)}$, that is one for each
\textquotedblleft would be Goldstone boson\textquotedblright . When $\mu
^{4}=m_{2}^{4}$, that is then model is at the zero exceptional point of type
I, the gauge mass vector bosons remain massless. This feature is compatible
with our previous observations in \cite{fring2019pseudo,fring2020goldstone}
and above, that at these points the Goldstone bosons can not be identified.

We notice here that the two massive vector gauge bosons are proportional to
the $\mathcal{CPT}$ -inner product of the symmetry broken vacuum solution%
\begin{equation}
m_{\text{gauge}}^{2}\propto \left\langle 0|0\right\rangle _{\mathcal{CPT}%
}\propto \phi ^{\text{vac}}\hat{I}\phi ^{\text{vac}}\propto R^{2}\left( 1-%
\frac{\mu ^{4}}{m_{2}^{4}}\right) .  \label{mgauge}
\end{equation}%
Hence the vanishing of the mass for the vector gauge bosons at the two types
of zero exceptional points can be associated to the vanishing of the $%
\mathcal{CPT}$ -inner product at these points. This is reminiscent of the
vanishing of the $\mathcal{CPT}$ -inner product at the standard exceptional
points, which is responsible for interesting phenomena such as the stopping
of light at these locations in the parameter space \cite%
{goldzak2018light,miri2019exceptional}. We note, however, a key difference
between the two scenarios: While the $\mathcal{CPT}$ -inner product in (\ref%
{mgauge}) is devised on the eigenvector space of squared mass matrix, the
latter is a $\mathcal{CPT}$ -inner product on the Hilbert space.

\section{Conclusions}

Employing a pseudo-Hermitian approach we found that the Higgs mechanism
applies in the usual way in the $\mathcal{CPT}$-symmetric regime by giving a
mass to the vector gauge bosons and preventing Goldstone bosons to exist,
which was also found in \cite{mannheim2018goldstone,alexanHiggs} using
different approaches. As in \cite{alexanHiggs} we also observed that in the
spontaneously broken $\mathcal{CPT}$-symmetric regime the vector gauge
bosons become massive and the Higgs mechanism is in tact. However, as in
this regime other particles acquire complex masses it has to be discarded as
non-physical for that reason. Even though technically one needs to treat the
standard exceptional point differently from the other regimes, the main
principle of the Higgs mechanism still holds up. In contrast to the finding
in \cite{alexanHiggs}, we observed that the Higgs mechanism breaks down at
the zero exceptional points, which was also observed in \cite%
{mannheim2018goldstone}. We find the same characteristic behaviour, i.e. the
matching of the amounts of massive vector gauge bosons and \textquotedblleft
would be Goldstone bosons\textquotedblright , for the complex scalar fields
taken in the fundamental as well as in the adjoint representation. The
vanishing of the mass for the vector gauge bosons coincides with the
vanishing of the $\mathcal{CPT}$ -inner product on the eigenvector space of
squared mass matrix.

Obviously there are many interesting extensions to these investigations,
such as for instance the treatment of models with a more involved field
content or different types of continuous symmetries.

\medskip

\noindent \textbf{Acknowledgments:} We would like to thank Alessandro de
Martino for useful comments.

\appendix\section{Type I (standard) versus type II (zero) exceptional points}

In this appendix we present a discussion that illustrates the difference between the two types of exceptional points. The main distinction in their behaviour is that a one-dimensional parameter space the $\mathcal{PT}$-symmetry is spontaneously broken on one side of the type I (standard) exceptional point, whereas for type II (zero) exceptional point the $\mathcal{PT}$-symmetry is preserved on both sides. The zero exceptional point occurs when two eigenvalues coalesce at zero, hence the name.

We consider here a ($3 \times 3$)-matrix of a very generic form that occurs for instance as a building block of the squared mass matrix in the model discussed in \cite{fring2020goldstone}, see equation (3.48) therein,  
\begin{equation}
H=\left(
\begin{array}{ccc}
	A & W & 0 \\
	-W & B & -V \\
	0 & V & -C \\
\end{array}
\right)
\end{equation}
Here we carry out the discussion for a Hamiltonian $H$, having in mind the analogy to the squared mass matrix. The determinant is easily computed to $\det H=A \kappa - C W^2$, $\kappa := V^2 -B C$. In order to obtain a zero eigenvalue, $\lambda_0=0$, we  enforce now the determinant to vanish by setting $A=W^2 C/\kappa$. The other two eigenvalues then become
\begin{equation}
\lambda_\pm=	\frac{\kappa  (B-C)+C W^2 \pm \tau}{2 \kappa } \label{A2}
\end{equation}
with $\tau=\sqrt{\kappa ^2 \left((B+C)^2-4 V^2\right)+2 \kappa  W^2 \left(C (B+C)-2 V^2\right)+C^2
W^4} $. 

The exceptional points are identified by simultaneously solving the two equations 
\begin{equation}
\det\left(H - \lambda \mathbb{I} \right)=0, \qquad \text{and} \qquad \frac{d}{d \lambda}\det\left(H - \lambda \mathbb{I} \right)=0, 
\end{equation}
for $W$ and $\lambda$, obtaining the two sets of eigenvalues
\begin{equation}
\lambda_\pm^e = \frac{\hat{\kappa}^2-\sqrt{\kappa } V}{C},\lambda_0=0 \qquad \text{and} \qquad 	\lambda_-^0 =\lambda_0=0,\lambda_+^0 =\frac{C^3+BV^2-2CV^2}{\hat{\kappa}^2} \label{A4}
\end{equation}
for the critical parameters 
\begin{equation}
W^e= \frac{\tilde{\kappa}}{C}, \qquad \text{and} \qquad W^0= i \frac{\kappa}{\hat{\kappa}}, \label{A5}
\end{equation}
respectively. We abbreviated $\tilde{\kappa}:= \pm\sqrt{\kappa  \left(\kappa +V^2-C^2\right) \pm 2 \kappa ^{3/2} V}$ and $\hat{\kappa}:= \sqrt{V^2-C^2}$. The first set of eigenvalues in (\ref{A4}) correspond to the standard exceptional point and the second set to the zero exceptional point.

Next we calculate the bi-orthonormal basis from the normalized left and right eigenvectors $u_i$, $v_i$, $i=0,\pm$, respectively, for $H$  
\begin{equation}
v_0 =\frac{1}{\sqrt{N_0}}(-\kappa, C W, V W), \quad v_\pm =\frac{1}{\sqrt{N_\pm}}(W(\hat{\kappa}^2 - C \lambda_\pm), \kappa(C+\lambda_\pm) , V \kappa), \quad  u_i= U v_i,
\end{equation}
with $U=diag(1,-1,1)$ and normalization constants $N_0= \kappa^2 + W^2 \hat{\kappa}^2$, $N_\pm=V^2 \kappa^2 + W^2 [ V^2-C(C+\lambda_\pm)]^2-(C+\lambda_\pm)^2 \kappa^2 $. By construction these vectors satisfy the orthonormality relation $u_i \cdot v_j = \delta_{ij}$.

We observe now that at the standard exceptional point the two eigenvectors for the non-normalised ($N_{\pm}$ become zero at the exceptional points) eigenvalues $\lambda_\pm^e$ coalesce, which distinguishes exceptional points from standard degeneracy. The left and right eigenvectors become in this case 
\begin{align}
v_\pm^{e,r}=\left( \tilde{\kappa}, V \sqrt{\kappa} -\kappa, C \sqrt{\kappa}   \right), \qquad & v_0^{e,r}=\left(- C \kappa, C \tilde{\kappa} , V \tilde{\kappa} \right), \\
v_\pm^{e,l}=\left( \tilde{\kappa}, \kappa-  V \sqrt{\kappa} , C \sqrt{\kappa}   \right),  \qquad & v_0^{e,l}=\left(- C \kappa, -C \tilde{\kappa} , V \tilde{\kappa} \right),
\end{align}%
with
\begin{equation}
v_\pm^{e,l} \cdot v_\pm^{e,r} = 0, \qquad \text{and} \qquad  v_0^{e,l} \cdot v_0^{e,r} = C^2\left( \kappa^2 - \tilde{\kappa}^2  \right) +V^2 \tilde{\kappa}^2 .
\end{equation}

Similarly, at the zero exceptional point the eigenvectors for the eigenvalues $\lambda_0$ and $\lambda_-^0 $ coalesce, which qualifies this point also to be called ``exceptional'' in the standard terminology. In this case the left and right eigenvectors become
\begin{align}
v_+^{0,r}=\left( V \kappa, i V \hat{\kappa}(C-B), i \hat{\kappa}^3   \right), \qquad & v^{0,r}=v_-^{0,r}=\left(i\hat{\kappa}, C , V  \right), \\
v_+^{0,l}=\left( V \kappa, i V \hat{\kappa}(B-C), i \hat{\kappa}^3   \right), \qquad & v^{0,l}=v_-^{0,r}=\left(i\hat{\kappa}, -C , V  \right),
\end{align}%
with
\begin{equation}
v_+^{e,l} \cdot v_+^{e,r} = \left( C^3 + B V^2 - 2 C V^2  \right)^2, \qquad \text{and} \qquad  v^{0,l} \cdot v^{0,r} = 0 .
\end{equation}

In order to understand the key difference between these two types of exceptional points we consider at first the eigenvalues (\ref{A2})  near the critical values in (\ref{A5}). Concerning the standard exceptional points we note that the two eigenvalues become identical when $\tau \rightarrow 0$. Thus writing $\tau/C^2= (W^2-(W^e)^2)(W^2-\tilde{W})$, with $\tilde{W}$ being the second root of the polynomial in $W^2$ under the square root, it is now clear that if we consider the eigenvalues as functions of $W^2$ the argument of the square root has different signs for $W^2=(W^e)^2 +\epsilon$ and $W^2=(W^e)^2 -\epsilon$. Hence the eigenvalues are real on one side of the exceptional point in the $W^2$-parameter space and complex on the other. In contrast none of the eigenvalues becomes complex in the neighbourhood of the critical value $W^0$. 

For completion we also report the Dyson map and hence the metric operator for which the same behaviour may be observed. Using the operator that diagonalises the non-Hermitian Hamiltonian $H$
\begin{equation}
\eta = (v_0,v_+,v_-), \qquad  \rho = \eta \eta^\dagger ,
\end{equation} 
with determinant
\begin{equation}
\det \eta = \frac{V \kappa}{\sqrt{N_0 N_+ N_-}} (\lambda_- -\lambda_+)  (\kappa^2 + W^2 \hat{\kappa}^2),
\end{equation} 
we verify the pseudo and quasi Hermiticity relations
\begin{equation}
\eta^{-1} H \eta = h=h^\dagger, \qquad   \rho H =H^\dagger \rho.
\end{equation}  
We observe that the map break down at both exceptional points, i.e. $\det \eta = 0$ for the critical values $W^e$ and $W^0$, and on one side of the standard exceptional point. In all other regions of the parameter space it holds. Thus we find the same behaviour as already observed for the analysis of the eigenvalues.

\newif\ifabfull\abfulltrue


\begin{thebibliography}{99}
\bibitem{englert1964broken} F.~Englert and R.~Brout, \newblock Broken
symmetry and the mass of gauge vector mesons, \newblock Phys. Rev. Lett. 
\textbf{13}(9), 321 (1964).

\bibitem{higgs1964brokena} P.~W. Higgs, \newblock Broken symmetries,
massless particles and gauge fields, \newblock Phys. Lett. \textbf{12},
132--133 (1964).

\bibitem{higgs1964brokenb} P.~W. Higgs, \newblock Broken symmetries and the
masses of gauge bosons, \newblock Phys. Rev. Lett. \textbf{13}(16), 508
(1964).

\bibitem{guralnik1964global} G.~S. Guralnik, C.~R. Hagen, and T.~W.~B.
Kibble, \newblock Global conservation laws and massless particles, \newblock %
Phys. Rev. Lett. \textbf{13}(20), 585 (1964).

\bibitem{Goddard:1986bp} P.~Goddard and D.~I. Olive, 
\newblock {Kac-Moody
and Virasoro Algebras in Relation to Quantum Physics}, \newblock Int. J.
Mod. Phys. \textbf{A1}, 303 (1986).

\bibitem{BPZ} A.~A. Belavin, A.~M. Polyakov, and A.~B. Zamolodchikov, %
\newblock Infinite conformal symmetry in two-dimensional quantum field
theory, \newblock Nucl. Phys. \textbf{B241}, 333--380 (1984).

\bibitem{YL1} C.-N. Yang and T.-D. Lee, \newblock Statistical theory of
equations of state and phase transitions. I. Theory of condensation, %
\newblock Phys. Rev. \textbf{87}(3), 404 (1952).

\bibitem{YL2} T.-D. Lee and C.-N. Yang, \newblock Statistical theory of
equations of state and phase transitions. II. Lattice gas and Ising model, %
\newblock Phys. Rev. \textbf{87}(3), 410 (1952).

\bibitem{fisher1978yang} M.~E. Fisher, \newblock Yang-Lee edge singularity
and $\phi$ 3 field theory, \newblock Phys. Rev. Lett. \textbf{40}(25), 1610
(1978).

\bibitem{cardy1985conformal} J.~L. Cardy, \newblock Conformal invariance and
the Yang-Lee edge singularity in two dimensions, \newblock Phys. Rev. Lett. 
\textbf{54}(13), 1354 (1985).

\bibitem{cardy1989s} J.~L. Cardy and G.~Mussardo, \newblock S-matrix of the
Yang-Lee edge singularity in two dimensions, \newblock Phys. Lett. B \textbf{%
225}(3), 275--278 (1989).

\bibitem{zamothermo} A.~B. Zamolodchikov, \newblock Thermodynamic Bethe
ansatz in relativistic models: Scaling 3-state Potts and Lee-Yang models, %
\newblock Nucl Phys. B \textbf{342}(3), 695--720 (1990).

\bibitem{zamo1991tri} A.~B. Zamolodchikov, \newblock From tricritical Ising
to critical Ising by thermodynamic Bethe ansatz, \newblock Nucl. Phys. B 
\textbf{358}(3), 524--546 (1991).

\bibitem{TBAHSG1999} O.~A. Castro-Alvaredo, A.~Fring, C.~Korff, and J.~L.
Miramontes, 
\newblock {Thermodynamic Bethe ansatz of the homogeneous
sine-Gordon models}, \newblock Nucl. Phys. \textbf{B575}, 535--560 (2000).

\bibitem{jacobsen2003} J.-L. Jacobsen, N.~Read, and H.~Saleur, \newblock %
Dense loops, supersymmetry, and Goldstone phases in two dimensions, %
\newblock Phys. Rev. Lett. \textbf{90}(9), 090601 (2003).

\bibitem{Geyer} H.~B. Geyer, F.~G. Scholtz, and I.~Snyman, \newblock %
Quasi-hermiticity and the Role of a Metric in Some Boson Hamiltonians, %
\newblock Czech. J. Phys. \textbf{54}, 1069--1073 (2004).

\bibitem{Bender:1998ke} C.~M. Bender and S.~Boettcher, \newblock Real
Spectra in Non-Hermitian Hamiltonians Having PT Symmetry, \newblock Phys.
Rev. Lett. \textbf{80}, 5243--5246 (1998).

\bibitem{Alirev} A.~Mostafazadeh, \newblock Pseudo-Hermitian Representation
of Quantum Mechanics, \newblock Int. J. Geom. Meth. Mod. Phys. \textbf{7},
1191--1306 (2010).

\bibitem{PTbook} C.~M. Bender, P.~E. Dorey, C.~Dunning, A.~Fring, D.~W.
Hook, H.~F. Jones, S.~Kuzhel, G.~Levai, and R.~Tateo, \newblock PT Symmetry:
In Quantum and Classical Physics, \newblock (World Scientific, Singapore)
(2019).

\bibitem{benderphi31} C.~M. Bender, V.~Branchina, and E.~Messina, \newblock %
Critical behavior of the P T-symmetric $i\phi^3$ quantum field theory, %
\newblock Phys. Rev. D \textbf{87}(8), 085029 (2013).

\bibitem{shalabyphi31} A.~M. Shalaby, \newblock Vacuum structure and P
T-symmetry breaking of the non-Hermetian $i\phi^3$ theory, \newblock Phys.
Rev. D \textbf{96}(2), 025015 (2017).

\bibitem{bender2018p} C.~M. Bender, N.~Hassanpour, S.~Klevansky, and
S.~Sarkar, \newblock PT-symmetric quantum field theory in D dimensions, %
\newblock Physical Review D \textbf{98}(12), 125003 (2018).

\bibitem{alexandre2015non} J.~Alexandre, C.~M. Bender, and P.~Millington, %
\newblock Non-Hermitian extension of gauge theories and implications for
neutrino physics, \newblock JHEP \textbf{2015}(11), 111 (2015).

\bibitem{rochev2015hermitian} V.~E. Rochev, \newblock Hermitian vs
PT-Symmetric Scalar Yukawa Model, \newblock arXiv preprint arXiv:1512.03286
(2015).

\bibitem{korchin2016Yuk} A.~Y. Korchin and V.~A. Kovalchuk, \newblock Decay
of the Higgs boson to $\tau^- \tau^+$ and non-Hermiticy of the Yukawa
interaction, \newblock Phys. Rev. D \textbf{94}(7), 076003 (2016).

\bibitem{laureyukawa} L.~Gouba, \newblock The Yukawa Model in One Space-One
Time Dimensions, \newblock in \emph{Mathematical Structures and Applications}%
, pages 225--233, Springer, 2018.

\bibitem{bender2005dual} C.~M. Bender, H.~F. Jones, and R.~J. Rivers, %
\newblock Dual PT-symmetric quantum field theories, \newblock Phys. Lett. B 
\textbf{625}(3-4), 333--340 (2005).

\bibitem{bender1999nonunitary} C.~M. Bender and K.~A. Milton, \newblock A
nonunitary version of massless quantum electrodynamics possessing a critical
point, \newblock J. of Phys. A: Math. and Gen. \textbf{32}(7), L87 (1999).

\bibitem{bender2000solution} C.~M. Bender, K.~A. Milton, and V.~M. Savage, %
\newblock Solution of Schwinger-Dyson equations for PT-symmetric quantum
field theory, \newblock Phys. Rev. D \textbf{62}(8), 085001 (2000).

\bibitem{milton2013pt} K.~A. Milton, E.~K. Abalo, P.~Parashar,
N.~Pourtolami, and J.~Wagner, \newblock PT-symmetric quantum electrodynamics
and unitarity, \newblock Phil. Trans. of the Royal Society A: Math., Phys.
and Eng. Sciences \textbf{371}(1989), 20120057 (2013).


\bibitem{Ohlsson} T. Ohlsson, \newblock Non-Hermitian neutrino oscillations
in matter with PT symmetric Hamiltonians, \newblock EPL \textbf{113,} 61001
(2016).

\bibitem{twoHiggs} G.C. Branco, P.M. Ferreira, L. Lavoura, M.N. Rebelo, M.
Sher and J.P. Silva, \newblock Theory and phenomenology of two-Higgs-doublet
models, \newblock Physics reports, \textbf{516}, 1-102 (2012).

\bibitem{multiHiggs} R.A. Flores and M. Sher, \newblock Higgs masses in the
standard, multi-Higgs and supersymmetric models, \newblock Annals of
Physics, \textbf{148}, 95-134 (1983).

\bibitem{schwinger51} J.~Schwinger, \newblock The theory of quantized
fields. I, \newblock Phys. Rev. \textbf{82}(6), 914 (1951).

\bibitem{alexandre2018spontaneous} J.~Alexandre, J.~Ellis, P.~Millington,
and D.~Seynaeve, \newblock Spontaneous symmetry breaking and the Goldstone
theorem in non-Hermitian field theories, \newblock Phys. Rev. D \textbf{98},
045001 (2018).

\bibitem{alex2019} J.~Alexandre, J.~Ellis, P.~Millington, and D.~Seynaeve, %
\newblock Spontaneously breaking non-Abelian gauge symmetry in non-Hermitian
field theories, \newblock Phys. Rev. D \textbf{101}(3), 035008 (2020).

\bibitem{mannheim2018goldstone} P.~D. Mannheim, \newblock Goldstone bosons
and the Englert-Brout-Higgs mechanism in non-Hermitian theories, \newblock %
Phys. Rev. D \textbf{99}(4), 045006 (2019).

\bibitem{fring2019pseudo} A.~Fring and T.~Taira, \newblock Pseudo-Hermitian
approach to Goldstone's theorem in non-Abelian non-Hermitian quantum field
theories, \newblock Phys. Rev. D \textbf{101}(4), 045014 (2020).

\bibitem{fring2020goldstone} A.~Fring and T.~Taira, \newblock Goldstone
bosons in different PT-regimes of non-Hermitian scalar quantum field
theories, \newblock Nucl. Phys. B \textbf{950}, 114834 (2020).

\bibitem{Dyson} F.~J. Dyson, \newblock Thermodynamic Behavior of an Ideal
Ferromagnet, \newblock Phys. Rev. \textbf{102}, 1230--1244 (1956).

\bibitem{alexanHiggs} J.~Alexandre, J.~Ellis, P.~Millington, and
D.~Seynaeve, \newblock Gauge invariance and the Englert-Brout-Higgs
mechanism in non-Hermitian field theories, \newblock Phys. Rev. D \textbf{99}%
(7), 075024 (2019).

\bibitem{Mostafazadeh:2002pd} A.~Mostafazadeh, \newblock Pseudo-Hermiticity
and Generalized PT- and CPT-Symmetries, \newblock J. Math. Phys. \textbf{44}%
, 974--989 (2003).

\bibitem{frith2019time} T.~Frith, \newblock Time-dependence in non-Hermitian
quantum systems, \newblock arXiv:2002.01977, PhD Thesis, City, University of
London (2019).

\bibitem{landau37} L.~D. Landau, \newblock On the theory of phase
transitions, \newblock Ukr. J. Phys. \textbf{11}, 19--32 (1937).

\bibitem{goldzak2018light} T.~Goldzak, A.~A. Mailybaev, and N.~Moiseyev, %
\newblock Light stops at exceptional points, \newblock Phys. Rev. Lett. 
\textbf{120}(1), 013901 (2018).

\bibitem{miri2019exceptional} M.-A. Miri and A.~Al{\`u}, \newblock %
Exceptional points in optics and photonics, \newblock Science \textbf{363}%
(6422), eaar7709 (2019).
\end{thebibliography}

\end{document}